\documentclass[12pt]{article}
\usepackage{amssymb,amsmath,amsthm,mathrsfs,latexsym,amsxtra,graphicx,appendix,epstopdf,feynmf,subfig,hyperref,setspace,fix-cm,color,cite,hyperref}
\numberwithin{equation}{section}
\onehalfspace
\begin{document}

\newcommand{\be}{\begin{equation}}
\newcommand{\ee}{\end{equation}}
\newcommand{\bea}{\begin{eqnarray}}
\newcommand{\eea}{\end{eqnarray}}
\newcommand{\bb}{\mathbb}
\newcommand{\mrm}{\mathrm}
\newcommand{\scr}{\mathscr}
\newcommand{\p}{\partial}
\def\e{{\rm e}}
\def\bz{{\bar z}}
\def\bw{{\bar w}}
\def\p{{$|\Phi\rangle$}}
\def\pp{{$|\Phi^\prime\rangle$}}
\def\cO{{\mathcal O}}
\def\cH{\mathcal{H}}
\def\cF{\mathcal{F}}
\def\cL{\mathcal{L}}

\newcommand{\CC}{\mathbb{C}} 
\newcommand{\ZZ}{\mathbb{Z}} 

\newcommand{\ie}{{\it i.e.~}}
\def\eg{{\it e.g.~}}

\newcommand{\comment}[1]{{\bf\color{blue}[#1]}}

\begin{titlepage}
\begin{center}

\hfill \\
\hfill \\
\vskip 0.75in

{\Large \bf Bootstrapping Chiral CFTs at Genus Two}\\

\vskip 0.4in

{\large Christoph A.~Keller${}^{a}$, Gr\'egoire Mathys$^{b}$, and Ida G.~Zadeh${}^{a}$
}\\
\vskip 4mm

${}^{a}$
{\it Department of Mathematics, ETH Zurich, CH-8092 Zurich, Switzerland} \vskip 1mm
${}^{b}$
{\it Department of Physics, ETH Zurich, CH-8092 Zurich, Switzerland} \vskip 1mm

\texttt{christoph.keller@math.ethz.ch, grmathys@student.ethz.ch,  zadeh@math.ethz.ch}

\end{center}

\vskip 0.35in

\begin{center} {\bf ABSTRACT } \end{center}
Genus two partition functions of 2d chiral conformal field theories are given by Siegel modular forms. We compute their conformal blocks and use them to perform the conformal bootstrap. The advantage of this approach is that it imposes crossing symmetry of an infinite family of four point functions and also modular invariance at the same time. Since for a fixed central charge the ring of Siegel modular forms is finite dimensional, we can perform this analytically. In this way we derive bounds on three point functions and on the spectrum of such theories.
\vfill

\noindent \today

\end{titlepage}

\setcounter{tocdepth}{2}

\tableofcontents

\section{Introduction}

Over the past few years there has been a lot of progress in the modern conformal bootstrap, started by \cite{Rattazzi:2008pe}. The modern conformal bootstrap combines crossing symmetry with numerical methods to obtain bounds on operator dimensions. With current technology, usually one checks crossing symmetry of a single four point function only. The ultimate goal is of course to bootstrap the entire CFT, for which it is necessary to combine all correlation functions and check for mutual consistency. In \cite{Kos:2014bka,Kos:2015mba,Kos:2016ysd} this was done for three four point functions, which lead to the 3d Ising model. For two dimensional CFTs, in addition to crossing symmetry we also require modular invariance. Similar to the crossing bootstrap, one can perform a modular bootstrap \cite{Hellerman:2009bu,Friedan:2013cba,Keller:2012mr,Collier:2016cls}, which gives another set of consistency conditions. The ultimate goal of the full bootstrap is to combine \emph{all} consistency conditions, and find solutions to them. For 2d CFTs in particular one wants to combine modular invariance and crossing symmetry at the same time. This is obviously a very hard problem.

Here we take a step towards this goal by combining an infinite number of constraints, both from crossing symmetry and from modular invariance. We do this by considering correlation functions on higher genus surfaces. Fundamentally, the idea behind this comes from the insight of \cite{Moore:1988qv}, namely that consistency of the correlation functions on the sphere, \ie crossing symmetry, and of the torus one-point functions, \ie modular invariance, implies that the higher genus correlation functions are well defined. In particular, they are invariant under the higher genus modular group. We can thus check an infinite number of crossing symmetry and modular invariance conditions at the same time.
Here we will consider the genus 2 partition function. In that case the modular group is $Sp(4,\ZZ)$. This group indeed contains the crossing group and the modular group. Using the technology of \cite{Gaberdiel:2009rd,Gaberdiel:2010jf}, we can relate the partition function to physical quantities such as three point functions of primaries and the spectrum of the theory.

As usual in the conformal bootstrap, there are two main ingredients. First, we need to compute the conformal blocks, in this case for $g=2$, $n=0$. To do this we choose Schottky coordinates on the moduli space, and obtain the conformal block using the technology of \cite{Gaberdiel:2009rd,Gaberdiel:2010jf}. The choice of Schottky coordinates allows us to translate the coefficients of the partition functions to three point functions. One could of course also use a different technology such as recursion relations to obtain the conformal block \cite{Cho:2017oxl}, although in that case one would still have to convert to Schottky coordinates, or some other suitable set of coordinates in order to connect to the appropriate physical correlators.

The second ingredient in the modern conformal bootstrap is to impose the symmetries. This is usually done by expanding around a fixed point of the symmetry and demanding that the first $N$ odd derivatives vanish. In our case we impose it instead by demanding that the genus 2 partition function be a Siegel modular form. This then automatically ensures that it has all the required symmetries. We will only consider chiral or meromorphic CFTs here, that is theories which only contain left-moving degrees of freedom. The advantage is then that for fixed central charge the space of such forms is finite dimensional. This means that we only need to fix a few physical parameters of the theory, and then automatically obtain all other physical quantities as functions of those parameters. The physical parameters that we fix are the spectrum of the lightest states and some of their three point functions. We will call this the `light data', $\cL$. The light data then automatically fixes the entire partition function, which in particular means that we can express an infinite number of (sums of squares of) three point functions in terms of it. Assuming unitarity and imposing that all these squares be positive then leads to an infinite family of constraints on the physical parameters, giving the sought-after bounds. Note that in this sense our bootstrap is not numerical, but analytic, as for instance in \cite{Bouwknegt:1988sv,Headrick:2015gba}.

Using this approach we investigate the allowed range of parameters for unitary meromorphic theories of central charge $c=8k$ with $c$ ranging from 8 to 72. The results for $c=24$ (and less) are of course known using a careful analysis of allowed spin 1 algebras \cite{Schellekens:1992db}. The results for higher central charges however are new. We find bounds on the number of states $N_h$ and the squares of three point functions. More precisely, let $c_{\varphi_1\varphi_2\varphi_3}$ be the three point function of three primary fields. The coefficients of the partition function are then given by
\be\label{C_hhh}
C_{h_1h_2h_3}=\sum_{\varphi \in \cH^{h_i}} c_{\varphi_1\varphi_2\varphi_3}c_{\varphi_3\varphi_2\varphi_1}
\ee
where the sum runs over all primary fields $\varphi_i$ of weight $h_i$. We can also define the average square of a three point function of primary fields of weights $h_1$, $h_2$, and $h_3$ by
\be\label{avg3pf}
\langle c_{h_1h_2h_3}^2\rangle = \frac{1}{N_{h_1}N_{h_2}N_{h_3}}C_{h_1h_2h_3}\ ,
\ee
where $N_h$ is the number of primary fields of weight $h$. These are the quantities that we will bound.

Bounds on three point functions using modular invariance or crossing symmetry were also considered for instance in \cite{Pappadopulo:2012jk,Kraus:2016nwo}. In those cases some of the operators were taken to be asymptotically heavy. In that sense our results are the analog results for light operators.

This paper is organized in the following way. In section \ref{s_W} we describe the ring of Siegel modular forms and their connection to the partition functions. In section \ref{s:confblock} we compute the conformal block of the genus 2 vacuum amplitude and in section~\ref{s_Z_q_p} we give explicit expressions for the genus 2 amplitudes in terms of the light data. In section~\ref{s:constraints} we derive constraints from unitarity.

\emph{Note added:} We coordinated submission with \cite{Alex} and \cite{Xi}, as there is some overlap in our results.

\section{Siegel Modular Forms}\label{s_W}

In this section we summarize some properties of Siegel modular forms which are important for our purposes, as well as their relation to partition functions. For a chiral CFT of central charge $c$, partition functions $Z_g$ can be written as a modular form $W_g$ of weight $c/2$ times some universal function $F_g$ which only depends on the genus \cite{Gaberdiel:2009rd,Gaberdiel:2010jf}
\be\label{Zg}
Z_g = \frac{W_g}{F_g^{c/2}}\ .
\ee
The function $F_g$ has the appropriate weight such that $Z_g$ is indeed invariant under modular transformations. In addition, the functions $W_g(\Omega)$, where $\Omega$ is the period matrix of the genus-$g$ Riemann surface, are holomorphic on the space of period matrices of the Riemann surface and are modular forms of degree $g$ and weight $k=c/2$:
\be\label{mf}
W_g\!\left(\frac{A\Omega+B}{C\Omega+D}\right)=\mathrm{det}(C\Omega+D)^k\,W_g(\Omega),\qquad
\bigg(\!\!\begin{array}{cc}A & B\\C & D \end{array}\!\!\bigg)\in\mathrm{Sp}(2g,\mathbb Z).
\ee
For $g\le3$, $W_g$ is a Siegel modular form. For $g=1$ this reduces to the definition of a usual modular form of weight $k$.

\subsection{Genus 1}
Here $W_1$ is an ordinary modular form transforming under $SL(2,\ZZ)$, and the Riemann period matrix is simply $\tau$. $F_1$ is given by the square of the $\eta$ function
\be\label{F1}
F_1 = \eta(\tau)^2,
\ee
where
\be
\eta= q^{\frac1{24}}\prod_{n=1}^\infty(1-q^n)\ .
\ee
The ring of modular forms is generated by the Eisenstein series
\be
G_4\ , \qquad G_6\ \label{eisen_g1_four},
\ee
where the subscript denotes the modular weight. For future use, we introduce also the modular discriminant of the elliptic curve, given by
\be\label{disc}
\Delta=\frac{G^3_4-G^2_6}{1728}=\eta(\tau)^{24} \ .
\ee

\subsection{Genus 2}\label{s_W_cons}
For genus 2 we parametrize the Riemann period matrix as
\be\label{Omega_g2}
\Omega=\bigg(\!\!\begin{array}{cc}\tau_{11}&\tau_{12}\\\tau_{21}&\tau_{22}\end{array}\!\!\bigg),
\ee
and the multiplicative periods as $q_{ij}=e^{2\pi i \tau_{ij}}$.
The generators of the ring of Siegel modular forms of degree 2  and even weight are \cite{igusa:1962}
\be\label{g2gens}
E_4,\qquad E_6,\qquad\chi_{10},\qquad\chi_{12}.
\ee
Here the $E_n$ are the genus 2 Eisenstein series, and $\chi_{10}$ and $\chi_{12}$ can be expressed in terms of Eisenstein series as we describe in appendix~\ref{s_eisen}. Following (\ref{disc}), we define the analogue genus 2 function
\be\label{psi12}
\psi_{12}=\frac{E^3_4-E^2_6}{1728},
\ee
which we will use in the upcoming sections.

As we will see in section~\ref{s:confblock}, in order to relate the genus 2 partition function to physical quantities, it is useful to go to Schottky space and work with Schottky coordinates $p_1$, $p_2$, and $x$,
\be
\mathfrak{S}_2 :=\{(p_1,p_2,x)\in\CC^3\mid x\neq 0,1,\ 0<|p_i|<\min\{|x|,\,\frac1{|x|}\}\ , i=1,2\}\ .
\ee
The relation between the multiplicative periods and the Schottky parameters are derived in appendix A of \cite{Gaberdiel:2010jf}. The power series expansion of $q_{ij}$ are of the form
	\bea
	&&q_{11}=p_1\sum_{n,m=0}^\infty\sum_{r=-n-m}^{n+m}c(n,m,|r|)\,p_1^np_2^mx^r,\label{q11ps}\\
	&&q_{22}=p_2\sum_{n,m=0}^\infty\sum_{r=-n-m}^{n+m}c(m,n,|r|)\,p_1^np_2^mx^r,\label{22ps}\\
	&&q_{12}=x+x\sum_{n,m=1}^\infty\sum_{r=-n-m}^{n+m}d(m,n,r)\,p_1^np_2^mx^r,\label{q12ps}
	\eea
	and $q_{21}=q_{12}$. The coefficients $c(n,m,|r|)$ and $d(n,m,r)$ are listed in appendix E of that paper.	
$F_2$ is essentially a generalization of the $\eta$ function to genus 2 and can be found in \cite{mcintyre}. We will only use its expansion in Schottky coordinates
\be\label{Fk2}
F_2=\sum_{n,m=0}^\infty\sum_{r=-n-m}^{n+m}b(n,m,|r|)\,p_1^n\,p_2^m\,x^r\ ,
\ee
with the coefficients $b$ given in \cite{Gaberdiel:2010jf}.

\subsection{Factorisation properties}\label{s_W_factor}
There are two constraints on the modular forms coming from their factorisation properties and from the action of the Siegel operator on them. They provide a way to relate higher genus partition functions to lower genus ones. These constraints have been studied in detail in \cite{Gaberdiel:2009rd} (see section 4.1 of this reference) and we summarise them below.

In the degeneration limit where a genus $g$ Riemann surface degenerates to a singular surface which has two smooth components of genus $g-k$ and $k$, the Riemann period matrix of the genus $g$ surface is block diagonal and the modular form $f_g$ factorises as
\be\label{factor_i}
f_g(\Omega^{(g)})\longrightarrow f_{g-k}(\Omega^{(g-k)})\otimes f_{k}(\Omega^{(k)}).
\ee
For $g=1$, assuming that the theory has a unique vacuum, this simply fixes the overall normalization as
\be\label{factor_vac}
\lim_{\tau\to i\infty}W_1(\tau)=1.
\ee
For genus 2, the factorisation properties of the generators of Siegel modular forms of degree 2 are
\be\label{genus2_factor}
E_4\to G_4\otimes G_4,\quad E_6\to G_6\otimes G_6,\quad\chi_{10}\to0,\quad\chi_{12}\to\Delta\otimes\Delta,
\ee
and for $\psi_{12}$ we have
\be\label{genus2_factor_psi12}
\psi_{12}\to G^3_4\otimes\Delta+\Delta\otimes G^3_4-1728\Delta\otimes\Delta.
\ee

Another constraint on Siegel modular forms comes from applying the Siegel operator on them. The Siegel operator is a linear map which maps a modular form of degree $g$ to one of degree $g-1$:
\be\label{factor_ii}
\Phi(f_g)=\Phi(f_1)\,f_{g-1}.
\ee
A cusp form of degree $g$ is any element of the kernel of this linear map. Using (\ref{factor_vac}), we find that
\be\label{factor_ii_W}
\Phi(W_g)=W_{g-1}.
\ee
For $g=1$ we have chosen the normalisation of degree one Eisenstein series in (\ref{eisen_g1_four}) such that the constant term in the Fourier expansion is 1. We then have
\be\label{genus1_sieg}
\Phi(G_4)=1,\qquad\Phi(G_6)=1.
\ee
The discriminant (\ref{disc}) is a cusp form of degree 1:
\be\label{genus1_sieg_cusp}
\Phi(\Delta)=0.
\ee
The action of the Siegel operator on the generators of modular forms of degree 2 is
\be\label{genus2_sieg}
\Phi(E_4)=G_4,\quad\Phi(E_6)=G_6,\quad\Phi(\chi_{10})=0,\quad\Phi(\chi_{12})=0,
\ee
and so $\chi_{10}$ and $\chi_{12}$ are cusp forms of degree 2. We also have
\be\label{genus2_sieg_psi12}
\Phi(\psi_{12})=\Delta.
\ee

\section{Genus two conformal block expansion}\label{s:confblock}
\subsection{Genus 2 partition function}
Let us now relate the partition functions to physical quantities. For the genus 1 partition function, this is of course straightforward: we simply have usual graded trace
\be
Z_1 = \sum_{h=0}^\infty \textrm{dim}\, \cH_h\,q^{h-c/24}\ .
\ee
The genus 2 partition function is a bit more subtle. We follow the approach of \cite{Gaberdiel:2010jf} and consider the surface to be a sphere of four punctures with two handles. In terms of the coordinates this means we perform a Schottky uniformisation of the genus 2 surface. The partition function is then given by a sum over four point functions on the sphere:
\be\label{partition_i}
Z_2=\sum_{h_1,h_2=0}^\infty C_{h_1,h_2}(x)\,p_1^{h_1}\,p_2^{h_2},
\ee
with $p_1$, $p_2$, and $x$ being the Schottky coordinates. Here there are 4 punctures on the sphere with cross ratio $x$, and two handles glue these punctures pairwise with coordinates $p_1$ and $p_2$ determining the shapes of these handles. The sum over functions $C_{h_1,h_2}(x)$ then represent the sum over four point functions on the sphere where fields of dimensions $h_1$ and $h_2$ run through the two handles:
\be\label{partition_ii}
C_{h_1,h_2}(x)=\sum_{\phi_i,\psi_i\in\mathcal H_{h_i}}G^{-1}_{\phi_1,\psi_1}G^{-1}_{\phi_2,\psi_2}
\Big\langle V^{out}(\psi_1,\infty)\;V^{out}(\psi_2,x)\;V^{in}(\phi_2,1)\;V^{in}(\phi_1,0)\Big\rangle\ ,
\ee
where $G_{\phi\psi}$ is the metric on the space of states. $C_{h_1,h_2}$ are almost the standard four point functions, except for a slightly different definition of the ``in" and ``out" vertex operators.

Let us explain this in more detail. First note that under a global conformal transformation $\gamma$, \ie a M\"obius transformation, a state transforms as
\be
U(\gamma)=\gamma'(z)^{L_0} e^{L_1\frac{\gamma''(z)}{2\gamma'(z)}}\ .
\ee
In particular, we have the transposition map $\hat \gamma : z \mapsto -\frac{1}{z}$, which gives \cite{Dolan:1994st,CFTBook}
\be
V(\phi,z)^T = V\left(z^{-2L_0} e^{-\frac{1}{z}L_1}\phi,-\frac1z\right)\ .
\ee
The metric $G_{\phi\psi}$ is then given by
\be
G_{\phi\psi} = \lim_{z\rightarrow0}\Big\langle V\left(z^{-2L_0} e^{-\frac{1}{z}L_1}\phi,-\frac1z\right)\;\;V(\psi,0)\Big\rangle = 
\lim_{z\rightarrow\infty}\langle V(z^{2L_0} e^{zL_1}\phi,z)\;\;V(\psi,0)\rangle\ .
\ee
Since we are interested in unitary theories, we want to introduce a hermitian structure. Let `` $\overline{\,\cdot\,}$ " be the antilinear involution which acts as hermitian conjugation on the Hilbert space. Note that it acts on operators as
\be
\overline{V_n(\psi)}= (-1)^{h+n}V_n(\bar\psi)\ .
\ee
We will usually choose our primaries to be real, that is to satisfy $\bar \varphi = \varphi$. Note that this choice implies for descendant fields that
\be
\bar \phi = (-1)^{N_\phi} \phi\ ,
\ee
where $N_\phi$ is the total level of all Virasoro descendants. Using this notation, the metric $G$ can be related to the standard Kac matrix $K_{\phi\psi}$ as
\be\label{Kac_i}
K_{\phi\psi}=G_{\bar\phi\psi} . 
\ee
It is useful to write (\ref{partition_ii}) as
\be\label{partition_iii}
C_{h_1,h_2}(x)=\sum_{\phi_i,\psi_i\in\mathcal H_{h_i}}K^{-1}_{\phi_1,\psi_1}K^{-1}_{\phi_2,\psi_2}
\Big\langle V^{out}(\bar\psi_1,\infty)\;V^{out}(\bar\psi_2,x)\;V^{in}(\phi_2,1)\;V^{in}(\phi_1,0)\Big\rangle\ .
\ee
The vertices $V^{in}$ and $V^{out}$ are defined as \cite{Gaberdiel:2010jf}
\bea\label{vinvout0inf}
&&\quad V^{in}(\phi_1,0) =\,V(\phi_1,0) = \phi_1(0)\ ,\label{vinvout0inf_i}\\
&&V^{out}(\psi_1,\infty)=\,V\bigg(U\Big(\hat\gamma(z)\Big)\psi_1,\infty\bigg)=\lim_{z\to\infty}V(z^{2L_0}\,e^{z\,L_1}\psi_1,z)\ ,\label{vinvout0inf_ii}
\eea
and

\bea\label{vinvout1x}
&&\;\,V^{in}(\phi_2,1)=V\Big((x-1)^{L_0}e^{-L_1}\phi_2,1\Big)\ , \label{vinvout1x_i}\\
&&V^{out}(\psi_2,x)=V\Big((x-1)^{L_0}e^{L_1}\psi_2,x\Big)\ ,\label{vinvout1x_ii}
\eea
and are the usual vertex operators for quasi-primary fields $\phi_i$ and $\psi_i$ since the corresponding states are annihilated by $L_{1}$. However, if $\phi_i$ and $\psi_i$ are not quasi-primary operators, then the action of $L_1$ on them is non-trivial and hence, the factors of $e^{L_1}$ are needed to be taken into account. Non-quasi-primary operators transform non-tensorially under M\"obius transformations and so acquire additional factors. These factors then render the four point functions crossing symmetric.

We take all primary fields to be orthonormal. We have
\bea
V^{in}(\phi_1,0)|0\rangle &=& |\phi_1\rangle,\\
\langle0|V^{out}(\bar\psi_1,\infty)&=& \langle\psi_1|\ .
\eea
Let us also define the corresponding three point functions. Namely,
\be\label{cout}
C^{out}(\phi_1,\phi_2,\phi_3;x)=x^{-h_3+h_1+h_2} \langle \phi_3|V^{out}(\bar\phi_2,x)|\phi_1\rangle\ 
\ee
and
\be\label{cin}
C^{in}(\phi_1,\phi_2,\phi_3;x)=\langle \phi_1|V^{in}(\phi_2,1)|\phi_3\rangle \ .
\ee
Inserting a complete set of states $\phi_3$, $\psi_3$, we then rewrite (\ref{partition_i}) as
\be\label{partition_v}
Z_2=\sum_{\phi_i,\psi_i\in\cH} 
\prod_{i=1}^3
K^{-1}_{\phi_i\psi_i}  C^{out}(\phi_1,\phi_2,\phi_3;x)C^{in}(\psi_1,\psi_2,\psi_3;x)\,p_1^{h_1}\,p_2^{h_2}\,x^{h_3-h_1-h_2}\ .
\ee
We shall also slightly rewrite the vertex operators (\ref{vinvout1x_i}) and (\ref{vinvout1x_ii}) to get nicer expressions for $C^{in}$ and $C^{out}$. Namely, we use
\be
V^{in}(\phi,1)=(x-1)^{h_\phi}V(e^{-\frac{1}{x-1}L_1}\phi,1)  
\ee
and
\be
V^{out}(\phi,x)= x^{L_0} x^{-h_\phi}(x-1)^{h_\phi} V(e^{\frac{x}{x-1} L_1}\phi,1) x^{-L_0}\ ,
\ee
to find
\bea\label{coutcin1}
C^{out}(\phi_1,\phi_2,\phi_3;x)&=&  (x-1)^{h_{\phi_2}}\langle \phi_3|V(e^{-\frac{x}{1-x} L_1}\bar\phi_2,1)|\phi_1\rangle\ , \\
C^{in}(\phi_1,\phi_2,\phi_3;x)&=& (x-1)^{h_{\phi_2}}\langle \phi_1|V(e^{-\frac{1}{x-1}L_1}\phi_2,1)|\phi_3\rangle\ .
\eea
These in fact have the property that 
\bea
&&C^{out}(\phi_1,\phi_2,\phi_3;x)^*=(x^*-1)^{h_{\phi_2}}\langle \phi_1|V(e^{-\frac{x}{1-x} L_1}\bar\phi_2,1)^\dagger|\phi_3\rangle\\
&&\qquad\qquad\qquad\qquad\;=(x^*-1)^{h_{\phi_2}}\langle \phi_1|V(\overline{e^{-L_1-\frac{x}{1-x} L_1}\bar\phi_2},1)|\phi_3\rangle\nonumber\\
&&\qquad\qquad\qquad\qquad\;=(x^*-1)^{h_{\phi_2}}\langle \phi_1|V(\overline{e^{-\frac{1}{1-x} L_1}\phi_2},1)|\phi_3\rangle\nonumber\\
&&\qquad\qquad\qquad\qquad\;=(x^*-1)^{h_{\phi_2}}\langle \phi_1|V(e^{-\frac{1}{x^*-1} L_1}\phi_2,1)|\phi_3\rangle=C^{in}(\phi_1,\phi_2,\phi_3;x^*)\ .\nonumber
\eea

\subsection{Genus two conformal block}\label{g2_conf_block}

We compute the genus 2 partition function (\ref{partition_v}) by decomposing it into conformal blocks:
\be\label{prttn_i}
Z_2(p_1,p_2,x)= \sum_{\varphi_1,\varphi_2,\varphi_3 \in \cH^p} c_{\varphi_1\varphi_2\varphi_3}c_{\bar\varphi_3\bar\varphi_2\bar\varphi_1} \cF_{2,0}(h_i,c;p_1,p_2,x)\ .
\ee
Here the triple sum is over all primary fields $\varphi$ of the theory, $c_{\varphi_1\varphi_2\varphi_3}$ is their three point function, and $\cF_{2,0}(h_i,c;p_1,p_2,x)$ is the conformal block for the genus 2 surface with zero punctures given in Schottky coordinates.

As usual in the conformal bootstrap, we now make the assumption that the theory is unitary. This means that the Kac matrix (\ref{Kac_i}) is positive definite, in particular also for primary fields. This means that we can choose an orthonormal basis for the primary fields. Note moreover that for any theory we can choose our primaries to be real, \emph{i.e.}, $\varphi = \bar\varphi$ \cite{CFTBook}. In this case the three point functions are either real or purely imaginary, depending on the weight of the fields, due to the general identity
\be
c_{\varphi_1\varphi_2\varphi_3}^*=(-1)^{h_1+h_2+h_3} c_{\bar\varphi_1\bar\varphi_2\bar\varphi_3}\ .
\ee
In addition, because of $c_{\varphi_1\varphi_2\varphi_3}=(-1)^{h_1+h_2+h_3}c_{\varphi_3\varphi_2\varphi_1}$ we have $c_{\varphi_1\varphi_2\varphi_3}c_{\bar\varphi_3\bar\varphi_2\bar\varphi_1}= |c_{\varphi_1\varphi_2\varphi_3}|^2$, and so the coefficients in front of the conformal block are necessarily non-negative. The Kac matrix is then simply the Kac matrix of the descendants, with the primaries being orthonormal.

For a primary field $\varphi$, we define the associated descendant state $|\varphi,\vec N\rangle$ as:
\be\label{desc_i}
|\varphi,\vec N\rangle = L_{N_1}L_{N_2}\cdots L_{N_n}|\varphi\rangle\ ,
\ee
where $\vec N = (N_1,N_2, \ldots N_n)$, and $-\vec N$ is a partition with $N_1 \leq N_2 \leq \ldots \leq N_n < 0$. We define $|\vec N|= N := \sum_i N_i$. The Kac matrix between two such states is then defined in the usual way,
\be\label{Kac}
K_{\vec N, \vec M}= \langle \varphi,\vec N| \varphi,\vec M\rangle\ .
\ee
Note that $K$ vanishes between two different primary fields, and only depends on $\varphi$ through its weight $h_\varphi$. For brevity, we have thus suppressed the dependence on $\varphi$. Using and (\ref{partition_v}) and (\ref{prttn_i}), we can then compute $\cF_{2,0}$:
\bea\label{g2block_i}
&&\cF_{2,0} =\\ 
&&|c_{\varphi_1\varphi_2\varphi_3}|^{-2}\sum_{\vec N_i, \vec M_i} \prod_{i=1}^3
K^{-1}_{\vec N_i,\vec M_i}  C^{in}(h_i,\vec M_i;x)C^{out}(h_i,\vec N_i;x) p_1^{h_1-N_1} p_2^{h_2-N_2} x^{h_3-N_3-h_1+N_1-h_2+N_2}\ .\nonumber
\eea

To compute explicit expressions, it is useful to define the `ordinary' three point function
\be\label{Cordinary}
C(h_i,\vec N_i)=\langle h_1,\vec N_1|V(|h_2,\vec N_2\rangle,1)|h_3,\vec N_3\rangle\ = \langle h_1,\vec N_1|V_{h_3-N_3-h_1+N_1}(|h_2,\vec N_2\rangle)|h_3,\vec N_3\rangle\ ,
\ee
where $V$ is the usual vertex operator (see section \ref{subs_3pfs} below for more details). Using this we shall then write $C^{in}$ and $C^{out}$ in terms of `ordinary' three point functions:
\bea\label{coutcin2}
C^{out}(h_i,\vec N_i;x)&=& (-1)^{N_2} (x-1)^{h_2-N_2}\sum_{l=0}^{-N_2}\frac{1}{l!} \left(\frac{-x}{1-x}\right)^{l}C(h_3,\vec{N}_3,h_2,L_1^l\vec N_2,h_1,\vec N_1)\ ,\nonumber\\
C^{in}(h_i,\vec N_i;x)&=&(x-1)^{h_2-N_2} \sum_{l=0}^{-N_2}\frac{1}{l!} (1-x)^{-l}C(h_1,\vec{N}_1,h_2,L_1^l\vec N_2,h_3,\vec N_3)\ .
\eea
Note  the extra factor of $(-1)^{N_2}$ in $C^{out}$ which accounts for the fact that in $C^{out}$ the operator inserted is $\bar\phi_2$ rather than $\phi_2$.

\subsection{Computing three point functions}\label{subs_3pfs}
Finally, let us explain how to compute the ordinary three point functions (\ref{Cordinary}) recursively for general descendant fields. We use the recursion formula of \cite{Gaberdiel:2000qn} to write the expression for the $\ell^{\mathrm{th}}$ descendant of the operator $\phi_\ell$ located at $x$ and $1$. This operator might be a Virasoro primary, the identity, or a descendant of them and so in general picking up the mode is non-trivial. In terms of the modes $V_n$, the general expression reads
\bea\label{GN_ii}
&&V_{-N_1}\Big(V_{-N_2}(\psi)\,\chi\Big)=\sum_{L\ge0}\Bigg({N_2+L-h_\psi\choose L}V_{-N_2-L}(\psi)\,V_{-N_1+N_2+L}(\chi)+\\
&&\qquad\qquad\qquad\qquad+(-1)^{N_2-h_\psi+2}{N_2+L-h_\psi\choose L}V_{-N_1+h_\psi-L-1}(\chi)\,V_{-h_\psi+L+1}(\psi)\Bigg)\ .\nonumber
\eea
In particular, for our purposes we have $\psi=T$ --- so that the modes $V_n(T)\equiv L_n$ are the Virasoro modes ---  and so the modes $V_n$ on the right hand side of equation (\ref{Cordinary}) are found to be given by
\bea\label{GN_iii}
&&V_{-N_1}\Big(L_{-N_2}\,\chi\Big)=\sum_{\ell\ge0}\Bigg({N_2+\ell-2\choose\ell}L_{-N_2-\ell}\,V_{-N_1+N_2+\ell}(\chi)+\\
&&\qquad\qquad\qquad\qquad+(-1)^{N_2}{N_2+\ell-2\choose\ell}V_{-N_1-\ell+1}(\chi)\,L_{\ell-1}\Bigg).\nonumber
\eea
We note that equation (\ref{GN_ii}) is valid for any value of the modes $N_1$, $N_2\in\mathbb Z$. For the non-negative values of the arguments of the binomial coefficients we have the standard expression
\be\label{binomstd}
{n\choose m}=\begin{cases}
\frac{n!}{m!(n-m)!}, & 0\le m<n,\\
0,&\text{otherwise}.
\end{cases}
\ee
For the negative values of the arguments, the extended binomial coefficients are given by
\be\label{binomext}
{n\choose m}=\begin{cases}
(-1)^m{-n+m-1\choose m}, & m\ge0,\\
(-1)^{n-m}{-m-1\choose n-m}, & m\le n,\\
0,&\text{otherwise},
\end{cases}
\ee
where $n<0$.

\section{Fixing partition functions}\label{s_Z_q_p}
\subsection{Conventions}\label{subs_conv}

As discussed in section~\ref{s_W}, partition functions can be expressed in terms of Siegel modular forms which have a finite-dimensional space. This means that for a fixed value of the central charge $c$, it is enough to fix finitely many parameters to obtain the full partition function. We will call this set of parameters the \emph{light data} $\mathcal{L}$. As we will see, the set $\cL$ consists on one hand of the multiplicities of the lightest fields in the theory, and on the other hand of a finite number of light three point functions.

Our conventions are the following: we denote by
\be
N_h,
\ee 
the number of \emph{primary} fields of weight $h$. We have, of course, $N_0=1$ which corresponds to the identity operator. We denote by
\be
c_{\varphi_1\varphi_2\varphi_3} \equiv \langle \varphi_1|\varphi_2(1)|\varphi_3\rangle,
\ee
the three point function of three primary fields $\varphi_i,\varphi_j,\varphi_k$, and by
\be\label{sum3pfsqrt}
C_{h_1h_2h_3}=\sum_{\varphi \in \cH^p_{h_i}} c_{\varphi_1\varphi_2\varphi_3}c_{\varphi_3\varphi_2\varphi_1}
=\sum_{\varphi \in \cH^p_{h_i}} |c_{\varphi_1\varphi_2\varphi_3}|^2\ ,
\ee
the sum over absolute squares of the three point functions of all primary fields of appropriate weight. We have of course $C_{hh0}= N_h$. The unitarity condition then reads
\be\label{unitarity}
C_{h_1h_2h_3}\geq 0\ ,
\ee
since each term in the sum $|c_{\varphi_1\varphi_2\varphi_3}|^2$ is a non-negative real number.

\subsection{Strategy}\label{ss:strategy}
The genus 1 partition function is completely fixed by $N_h$. More precisely, it is fixed by the number of primary fields with dimension $h$,
\be\label{Nlight}
h\in\{0,\ldots, \lfloor \frac{c}{24} \rfloor\}\ ,
\ee
where $\lfloor~\rfloor$ denotes the floor function. This means that all the higher $N_h$ with $h>\lfloor\textstyle\frac{c}{24}\rfloor$ can be expressed as linear functions of the first few $N_h$. The genus 2 partition function is of the form
\be\label{Z_q}
Z_2=\frac{W_2}{F^{k}_2}\ ,
\ee
where $c=2k$. As discussed in subsection \ref{s_W_cons}, for fixed central charge the space of $W_2$ is finite dimensional, so that there are only a finite number of free parameters. Our goal is again to fix these parameters in terms of the light data $\cL$. To do this we proceed in two steps.

In a first step, we use the factorisation properties of the modular forms as well as the action of the Siegel operator on these forms as described in section~\ref{s_W_factor} to relate it to the genus 1 partition function. This fixes some of the parameters in terms of the $N_h$. For $c\leq24$, we will see that this completely fixes all the coefficients. The light data thus only consists of the $N_h$ given in (\ref{Nlight}). For $c>24$, however, there are still free parameters after this procedure: the partition function is thus not uniquely fixed by the spectrum of the theory, but also depends on a \emph{finite} number of the three point function coefficients. In these cases, we have to perform a second step: we write the partition function $Z_2$ in terms of the conformal blocks as in (\ref{prttn_i}) with general coefficients $C_{h_1h_2h_3}$ (see equation (\ref{g2block_i})). We then compare this with the expansion of (\ref{Z_q}) in Schottky coordinates, matching the coefficients term by term. To fix the remaining free parameters of $W_2$, is then enough to specify just a finite number of the lightest $C_{h_1h_2h_3}$. These are the ones that comprise the second part of the the light data $\cL$. We shall now carry out these steps and compute the partition functions of CFTs for different values of the central charge. The results are outlined in the next subsection.

\subsection{Expressions}

\subsubsection{c=8}\label{subsubs_g1_8}
The only degree 1 modular form of weight $k=4$ is the Eisenstein series $G_4$. Equations (\ref{factor_ii_W}) and (\ref{genus1_sieg}) then fix the normalization to give
\be\label{Wg1c8_ii}
W^{c=8}_1=G_4\ .
\ee
Similarly the genus 2 form is given by the weight $4$ form $E_4$, \be\label{Wg2c8_ii}
W^{c=8}_2=E_4\ ,
\ee
where, using (\ref{factor_ii_W}) and (\ref{genus2_sieg}), the overall constant is fixed to 1. We then have
\be\label{Z_g2_c8}
Z_2^{c=8}=\frac{E_4}{\left(F^{12}_2\right)^{\frac13}}\ ,
\ee
where $F^{k}_2$ is the reference partition function defined in (\ref{Fk2}). This describes the $E_8$ lattice CFT.

\subsubsection{c=16}\label{subsubs_g1_16}
Similar to the previous case, the only contribution to the modular forms $W_1$ of weight $k=8$ comes from $G_4$:
\be\label{Wg1c16}
W^{c=16}_1=G_4^2\ .
\ee
For $c=2k=16$ we find
\be\label{Wg2c16}
W^{c=16}_2=E_4^2,\qquad\qquad Z_2^{c=16}=\frac{E_4^2}{\left(F^{12}_2\right)^{\frac23}}\ .
\ee
This is either the $E_8\times E_8$ or the $SO(32)$ lattice theory. Note that up to genus 2, we cannot see the difference between the two.

\subsubsection{c=24}\label{subsubs_g1_24}
This is the first case where we have a free parameter, as we now have $G_4^3$ and $G_6^2$, or equivalently $G_4^3$ and $\Delta$, contributing to $W_1$:
\be\label{Wg1c24_i}
W^{c=24}_1=a_1G_4^3+a_2\Delta\ .
\ee
Using the constraint (\ref{factor_ii_W}), we find that $a_1=1$ and so
\be\label{Wg1c24_ii}
W^{c=24}_1=G_4^3+a\Delta\ .
\ee
To fix $a$, we write the genus one chiral character of the CFT which is the form  \cite{Gaberdiel:2009rd}:
\be\label{g1char}
\chi_{1}^{c}=\left(\frac{q}{\eta^{24}}\right)^\frac{c}{24}W_1^c\ .
\ee
Note that we have shifted the definition of the partition function (\ref{Zg}) by an overall factor of $q^{c/12}$ for the leading term in the expansion to be 1. For the CFT with $c=24$ we then find
\be\label{g1charc24}
\chi_{1}^{c=24}=q\,\frac{(G_4^3+a\Delta)}{\Delta}=q(j(\tau)+a)=1+(744+a)q+196884q^2+21493760q^3+O(q^4)\ ,
\ee
where we have used equation (\ref{disc}). This lets us identify the coefficient of $q$ with the number of spin 1 currents:
\be\label{g1charc24_ii}
N_1=744+a\ ,
\ee
thus fixing the only free parameter $a$ in terms of the light data $\cL=\{N_1\}$:
\be\label{Wg1c16_ii}
W^{c=24}_1=G_4^3+(N_1-744)\Delta\ .
\ee
For the genus 2 amplitude there are three possible contributions:
\be\label{Wg2c24}
W^{c=24}_2=b_1E_4^3+b_2\psi_{12}+b_3\chi_{12}\ .
\ee
Imposing the two constrains coming from applying the Siegel operator and the factorisation property of $W_2$ as in section~\ref{s_W_factor} fixes all these three free parameters as
\be
b_1=1,\qquad b_2=N_1-744,\qquad b_3=(N_1-744)(N_1+984)\ .
\ee
We then have
\be\label{Wg2c24_iii}
Z_2^{c=24}=\frac{W^{c=24}_2}{F^{12}_2},
\ee
where
\be\label{Z_g2_c24}
W^{c=24}_2=E_4^3+(N_1-744)\psi_{12}+(N_1-744)(N_1+984)\chi_{12}.
\ee
The genus 1 and 2 partition functions are completely fixed in terms of $N_1$.

\subsubsection{c=32}\label{subsubs_g1_32}
This is the first case where the light data contains multiplicities as well as three point functions, namely, $\cL=\{N_1,C_{111}\}$. $W_1$ is given by
\be\label{Wg1c32}
W^{c=32}_1=G_4\,\Big(G_4^3+(N_1-992)\Delta\Big)\ ,
\ee
and the genus 2 amplitude reads
\bea\label{Z_g2_c32}
&&Z_2^{c=32}=\frac1{\left(F^{12}_2\right)^{\frac43}}\Big(E_4^4+(N_1-992)\,E_4\,\psi_{12}+(N_1-992)(N_1+736)\,E_4\,\chi_{12}+\nonumber\\
&&\qquad\qquad\qquad\qquad\quad+\,\frac13\, (N_1^2-520N_1+246016 -12C_{111})\,E_6\,\chi_{10}\Big).
\eea
We note that the coefficient of the cusp form $\chi_{10}$ is not fixed by the constraint equations from factorisation (\ref{genus2_factor}) and Siegel operator (\ref{genus2_sieg}). To find this coefficient, we expand the partition function in terms of the Schottky coordinates, and match it with the conformal block expansion of the partition function described in subsection \ref{g2_conf_block}. We then evaluate the coefficient of the cusp form in terms of $N_1$ and $C_{111}$.

\subsubsection{c=40}\label{subsubs_g1_40}
Here the light data is $\cL=\{N_1, C_{111},C_{222}\}$. The modular form $W_1^{c=40}$ is given by
\be\label{Wg1c40_ii}
W^{c=40}_1=G_4^2\,\Big(G_4^3+(N_1-1240)\Delta\Big),
\ee
and the genus 1 partition function is found to be
\bea\label{g1charc40}
&&\chi_{1}^{c=40}=q\Big(j(\tau)+(N_1-1240)\Big)\chi_1^{c=16}\nonumber\\
&&\qquad\;\;=1+N_1q+(496N_1+20620)q^2 +(69752N_1+86666240)q^3+O(q^4).
\eea
The genus 2 partition function is of the form
\bea\label{Z_g2_c40}
Z_2^{c=40}=&&\!\!\!\!\!\!\!\!\frac1{\left(F^{12}_2\right)^{\frac53}}\bigg(E_4^5+(N_1-1240)E_4^2\,\psi_{12}+(N_1-1240)(N_1+488)\,E_4^2\,\chi_{12}+\\
&&\qquad\qquad+\,\frac13\, (N_1^2-1016N_1+615040 -12C_{111})\,E_4\,E_6\,\chi_{10}+\nonumber\\
&&\quad\;+\Big(9312 N_1^2 - 20665776 N_1 + \frac{77050410528}{5} - 905520 C_{111} + 16 C_{222}  \Big)\,\chi_{10}^2\bigg),\nonumber
\eea
which is expressed in terms of $N_1$, $C_{111}$, and $C_{222}$.

\subsubsection{c=48}\label{subsubs_g1_48}\label{s_Z_c48}
The light data is now $\cL=\{N_1,N_2,C_{111}, C_{222}\}$. We have
\be\label{Wg1c48_ii}
W^{c=48}_1=G_4^6+(N_1-1488)G_4^3\Delta+(N_2-743N_1+159769)\Delta^2\ .
\ee
The genus 1 partition function reads
\bea\label{g1charc48}
&&\chi_{1}^{c=48}=q^2\,\Big(j(\tau)^2+(N_1-1488)j(\tau)+(N_2-743 N_1+159769)\Big)\nonumber\\
&&\qquad\;\;=1+N_1q+(1+N_1+N_2)q^2 +(196884N_1+42987520)q^3+\nonumber\\
&&\qquad\qquad\;\,+\,(21493760N_1+40491909396) q^4+\nonumber\\
&&\qquad\qquad\;\,+\,(842806210 N_1+8463554690796) q^5+O(q^6)\ ,
\eea
and is determined in terms of $N_1$ and $N_2$. The higher values of $N_h$, $h>2$, are then fixed in terms of $N_1$ and $N_2$. The genus 2 amplitude reads
\bea\label{Z_g2_c48}
Z_2^{c=48}=&&\!\!\!\!\!\!\!\!\frac1{\left(F^{12}_2\right)^{2}}\bigg(E_4^6+\big(N_1-1488\big)E_4^3\,\psi_{12}+\big(N_1^2+ 238 N_1-2 N_2-676658\big)E_4^3\,\chi_{12}+\nonumber\\
&&\qquad\quad\;\;\,+\,\big(-743 N_1 + N_2+159769\big)\psi_{12}^2+\big(-731855 N_1^2 - 269472862 N_1+\nonumber\\
&&\qquad\quad\;\;\;+\, 242 N_1 N_2 +N_2^2+ 734258 N_2 +91785533041\big)\chi_{12}^2+\nonumber\\
&&\qquad\quad\;\;\,-\,\big(N_1+1968) (743 N1-N2-159769\big)\psi_{12}\,\chi_{12}+\nonumber\\
&&\qquad\quad\;\;\,+\,\frac13\, \big(N_1^2-26 N_1-2 N_2+787534 -12C_{111}\big)E^2_4\,E_6\,\chi_{10}+\nonumber\\
&&\qquad\quad\;\;\; +\,\frac{1}{9}\, \big(819977 N_1^2+137560792 N_1-392 N_1 N_2-N_2^2-855434 N_2+\nonumber\\
&&\qquad\quad\;\;\; - \, 10238832 C_{111}+144 C_{222}+229938936071\big)\, E_4\,\chi_{10}^2\bigg)\ ,
\eea
and is completely determined in terms of $N_1$, $N_2$, $C_{111}$, and $C_{222}$, as expected. 

A special case of the our expressions correspond to the proposed extremal CFTs \cite{Witten:2007kt}. These theories have no primary fields with dimension $h\le c/24$ and so the three point functions including these primaries vanish. For the case of $c=48$, this simply means that $N_1=N_2=0$, and therefore also $C_{111}=C_{222}=0$, so then we can immediately obtain the genus 2 partition function from (\ref{Z_g2_c48}). Note that this was already computed in  \cite{Gaiotto:2007xh}.

\vskip 4mm

It is clear that we can continue this procedure to higher central charge, the only difference being that the expressions will become more and more complicated. We have also computed the genus 2 amplitude for theories with $c=56$, $64$, and $72$. The full expressions are given in appendix \ref{app_Z2}.

\section{Constraints}\label{s:constraints}
From the expressions we obtained in the previous section for genus 2 amplitudes, and from conformal block expansion described in section \ref{g2_conf_block}, we can read off an infinite number of identities expressing an infinite number of three point functions and the total number of primary fields in terms of finitely many free parameters of the light data:
\be
N_h(\cL)\ , \qquad C_{h_1h_2h_3}(\cL)\ .
\ee
This constrains the parameter space, \emph{i.e.}, the allowed light data $\cL$, since we necessarily have
\be\label{Nh_bound}
N_h \geq 0\ ,
\ee
and, for unitary theories, we have
\be\label{3pf_bound}
C_{h_1h_2h_3}\geq 0\ .
\ee
We can then find bounds on the squares of three point functions $C_{h_1h_2h_3}$ (\ref{sum3pfsqrt}), and also on averages of squares of three point functions defined as
\be\label{3pf_avg_bound}
\langle c_{h_1h_2h_3}^2\rangle = \frac{1}{N_{h_1}N_{h_2}N_{h_3}}C_{h_1h_2h_3}\ .
\ee
We will next discuss constraints we find for each value of the central charge.

\subsection{c=24}
The only free parameter is the number of currents $N_1$. Using the genus 1 partition function, we can read off the expressions for the $N_h$,
\bea
&&N_2 = 196883 - N_1\ ,\qquad N_3=21296876 - N_1\ ,\qquad\qquad N_4 = 842609326\ ,\\
&&N_5 = 19360062527\ ,\qquad N_6=312092484374 + N_1\ ,\qquad N_7=3898575000125\ .\nonumber
\eea
Equation (\ref{Nh_bound}) then already gives a (weak) upper bound on $N_1$, namely $N_1 \leq 196883$.
To improve on this bound, we turn to the coefficients $C_{h_1h_2h_3}$. Starting with lowest weights,
 \be\label{c24_C111_ii}
 C_{111}=\frac1{12}N_1(N_1-24),
 \ee
we find that the condition (\ref{3pf_bound}) gives
\be\label{c24_N1_i}
N_1=0,\qquad N_1\ge24\ ,
\ee
which gives a lower bound.
The next highest coefficient is
\be\label{c24_C112_ii}
C_{112}= C_{121}= C_{211}=\frac{23}{24}N_1(N_1+2),
\ee
which gives $N_1\ge0$ and so does not improve the bound (\ref{c24_N1_i}). We note that the sums over the square of the three point function coefficients are symmetric under the exchange of the indices, as expected. We next have
\be\label{c24_C122_iii}
C_{122}=\frac{5}{12}N_1(78744-5N_1),
\ee
which introduces a tighter upper bound on the number of currents
\be\label{c24_N1_ii}
0\le N_1\le 15748.
\ee
Expanding the partition function \eqref{Z_g2_c24} up to order $O(p_1^5\,p_2^5)$,  we find that the coefficient which yields the stringent upper bound is
\be\label{c24_C223_iii}
C_{223}=12011438064-\frac{1}{468} N_1 \left(107015 N_1+38610024\right),
\ee
giving
\be\label{c24_N1_iii}
0\le N_1\le 7059.
\ee
All in all, we find
\be\label{c24_N1_iv}
N_1=0,\quad\mathrm{or}\quad24\le N_1\le 7059\ .
\ee
We note that this bound is less restrictive than the one obtained by Schellekens \cite{Schellekens:1992db}, \emph{i.e.}, $N_1=0$ or $24\leq N_1\leq 1128$. We could of course improve our bounds by following his analysis of the allowed spin 1 algebras. In the spirit of our bootstrap approach we do not do this, since in particular it would not generalize to constraining higher spin fields.

\subsection{c=32}

\begin{figure*}
    \centering
    \begin{minipage}[b]{.45\linewidth}
        \centering
        \includegraphics[width=7cm]{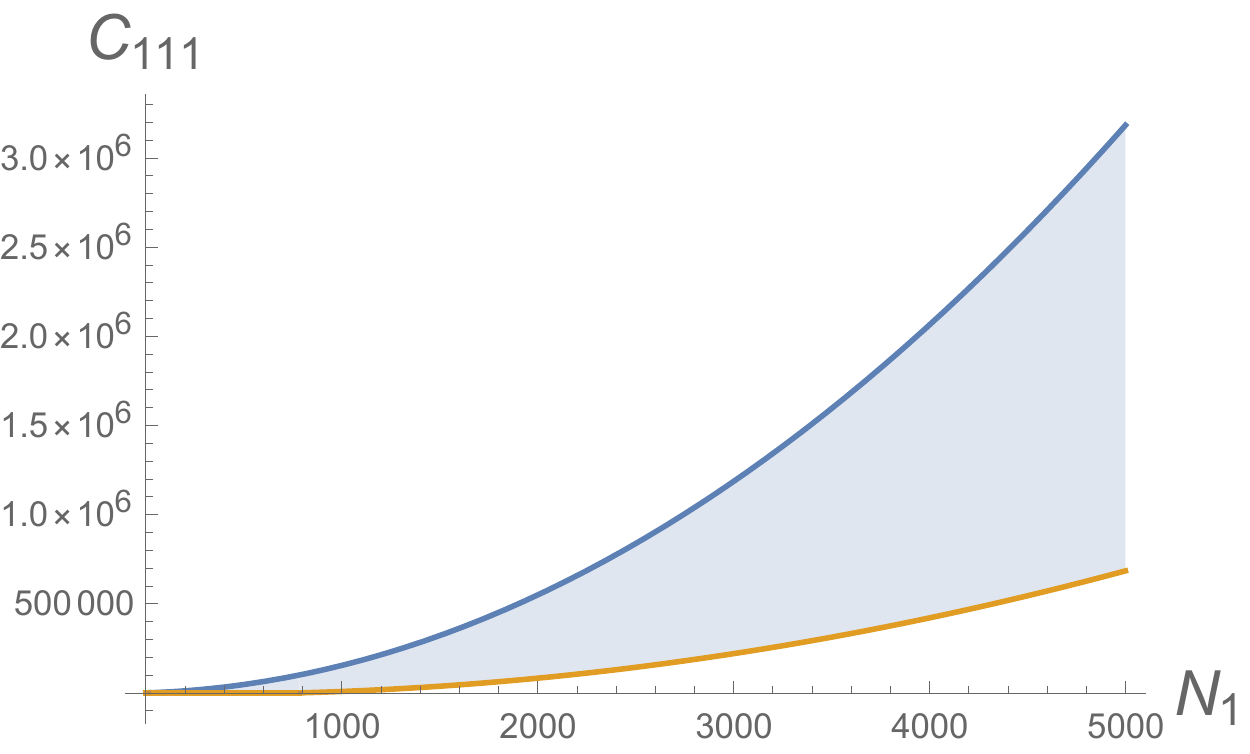}
    \end{minipage}%
    \hfill%
    \begin{minipage}[b]{.45\linewidth}
    \centering
        \centering
        \includegraphics[width=7cm]{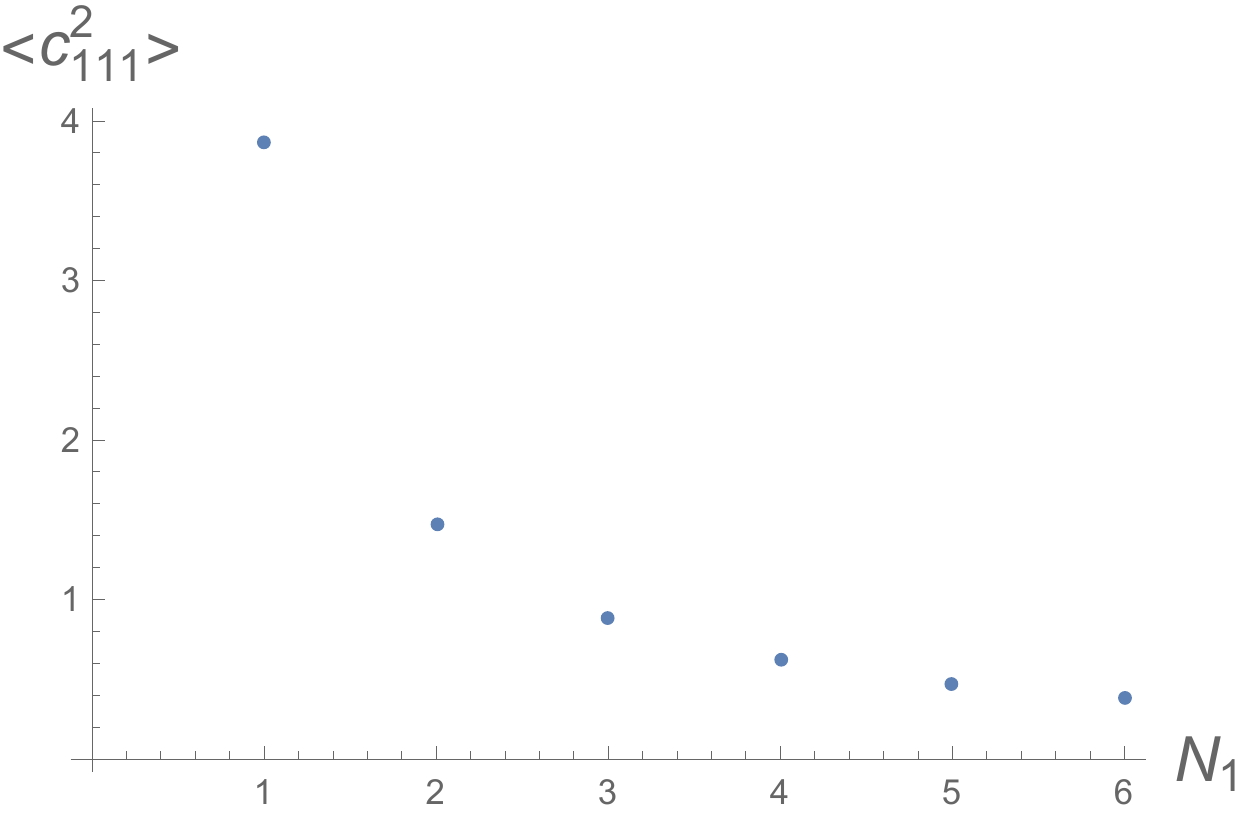}
    \end{minipage}\\[-7pt]
    \begin{minipage}[t]{.45\linewidth}
        \caption{Upper and lower bounds on $C_{111}$ for the $c=32$ theory. }
       \label{c32_1}
    \end{minipage}%
    \hfill%
    \begin{minipage}[t]{.45\linewidth}
        \caption{Upper bound on $\langle c_{111}^2\rangle=\textstyle\frac{C_{111}}{N_1^3}$ for $c=32$.}
        \label{c32_3}
    \end{minipage}%
\end{figure*}

The genus 1 partition function of the $c=32$ theory is fixed by $N_1$. The multiplicities $N_h$ for the first few low lying values of $h$ are given by
\begin{align}
N_2 &= 13 \left(19 N_1+10731\right),& N_3 &= 3875 N_1+69193488,\label{numberstatesc32_1}\\
N_4 &=30380 N_1+6928824200, & N_5&= 174250 N_1+322955200393.\label{numberstatesc32_2}\nonumber
\end{align}

Turning to the genus 2 partition function (\ref{Z_g2_c32}) and imposing positivity (\ref{3pf_bound}) for the higher $C_{h_1h_2h_3}$, we obtain the following upper and lower bounds for $C_{111}$:
\be
C_{111}\leq
\begin{cases}
	\frac{1}{8} \left(16 N_1^2+15 N_1\right) & 0\leq N_1<
	17, \\
	\frac{1}{497} \left(60 N_1^2+16440 N_1\right) &
	17\le N_1< 97359, \\
	\frac{1}{367} \left(21 N_1^2+2281170 N_1\right) &
	97359\le N_1< 4118426, \\
	\frac1{204575} (9900 N_1^2+8709154344N_1)&
	N_1\ge4118426.
\end{cases}
\label{C111maxc32}
\ee
and
\be
C_{111}\geq
\begin{cases}
	0 & 0\leq N_1< 787, \\
	\frac1{56116} (1768 N_1^2-1119161 N_1-213439590)&
	N_1\ge787.
\end{cases}
\label{C111minc32}
\ee
Figure~\ref{c32_1} shows the allowed region for $C_{111}$. The bounds obtained in (\ref{C111maxc32}) and (\ref{C111minc32}) grow quadratically in $N_1$.

It is instructive to consider the averaged quantity $\langle c_{111}^2\rangle = \textstyle\frac{C_{111}}{N_1^3}$ defined in (\ref{avg3pf}). Figure~\ref{c32_3} shows the upper bound on $\langle c_{111}^2\rangle$ as a function of $N_1$. Maximizing over $N_1$, we find a global upper bound
\be
\langle c_{111}^2\rangle =\frac{C_{111}}{N_1^3}\leq \frac{31}{8}.
\label{avgC111c32}
\ee
Using the bounds obtained in (\ref{C111maxc32}) and (\ref{C111minc32}), we can also obtain global bounds on the averages of the squares of higher three point functions $\langle c^2_{h_1h_2h_3}\rangle$. For the first few low lying coefficients we find
\be 
\langle c_{112}^2\rangle \le 1.3864 \cdot 10^{-5},\qquad\langle c_{222}^2\rangle \leq 1.22274\cdot 10^{-7}\ .
\label{avgc112c32}
\ee
The fact that these three point functions are so small on average indicates that there is some symmetry leading to a selection rule. On the other hand, for $C_{333}$ the upper bound has its maximum at $N_1 = 97359$, giving a global bound
\be 
\langle c_{333}^2\rangle \leq 0.31\ ,
\ee
which is much larger than (\ref{avgc112c32}).

\subsection{c=40}

\begin{figure}[t]
	\begin{center}
		\includegraphics[scale=0.5]{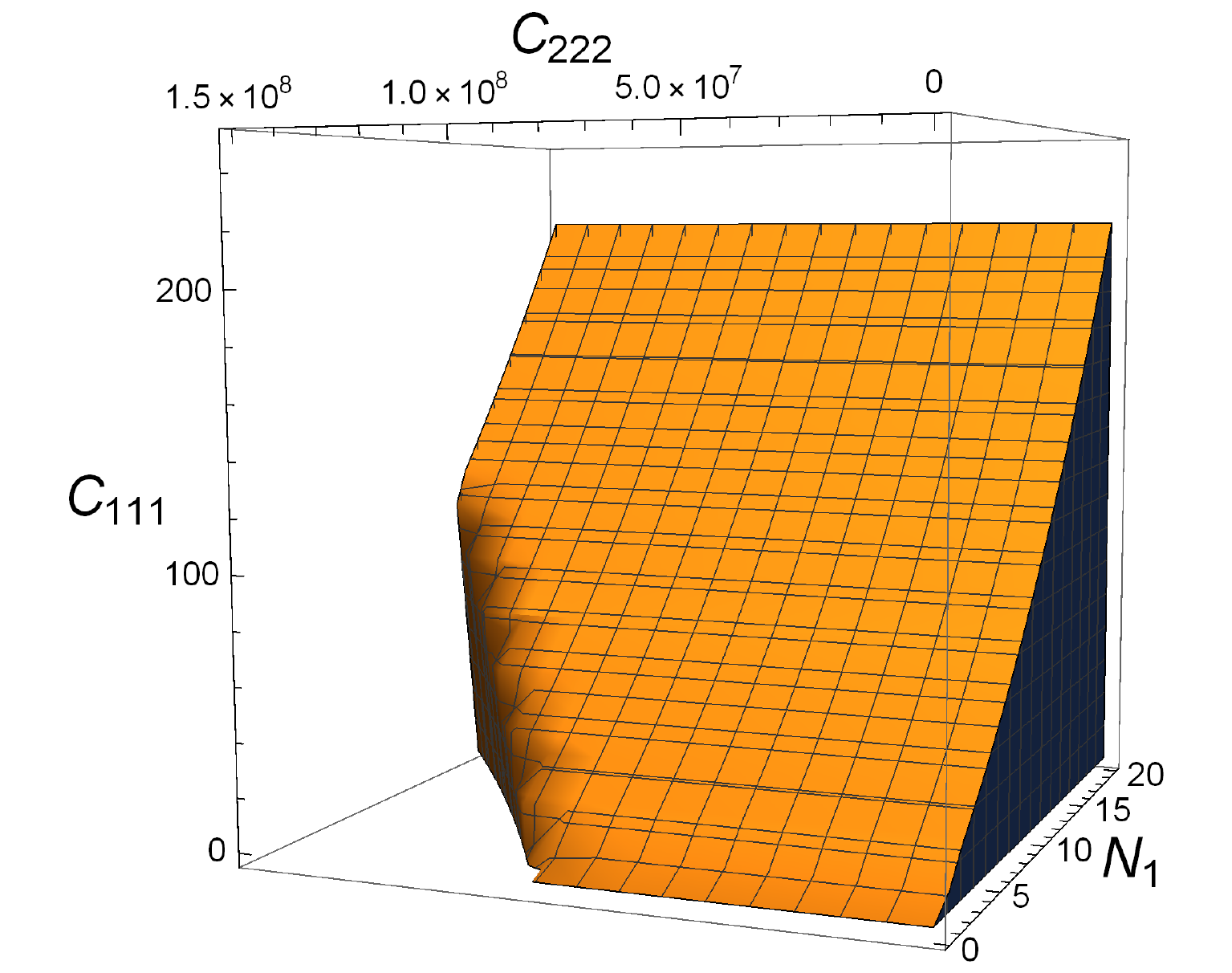}
		\caption{The allowed region for the light data $N_1$, $C_{111}$, and $C_{222}$, satisfying the unitarity constraints for the $c=40$ theory.}
		\label{c40_3d}
	\end{center}
\end{figure}

The genus one partition function is still determined by $N_1$ alone, giving
\begin{align}
N_2 &= 9 \left(55 N_1+2291\right),& N_3 &= 69255 N_1+86645620,\label{numberstatesc40_1}\\
N_4 &=2044760 N_1+24157197490, & N_5&= 81 \left(401060 N_1+28684566739\right).\nonumber\label{numberstatesc40_2}
\end{align}
The genus 2 partition function now depends on the light data $N_1$, $C_{111}$ and $C_{222}$. Imposing the constraints (\ref{Nh_bound}) and (\ref{3pf_bound}), we can determine the  allowed region for these three parameters, which is depicted in figure~\ref{c40_3d}.

\begin{figure}[t]
	\centering
	\begin{minipage}{0.5\textwidth}
		\centering
		\includegraphics[width=1\textwidth]{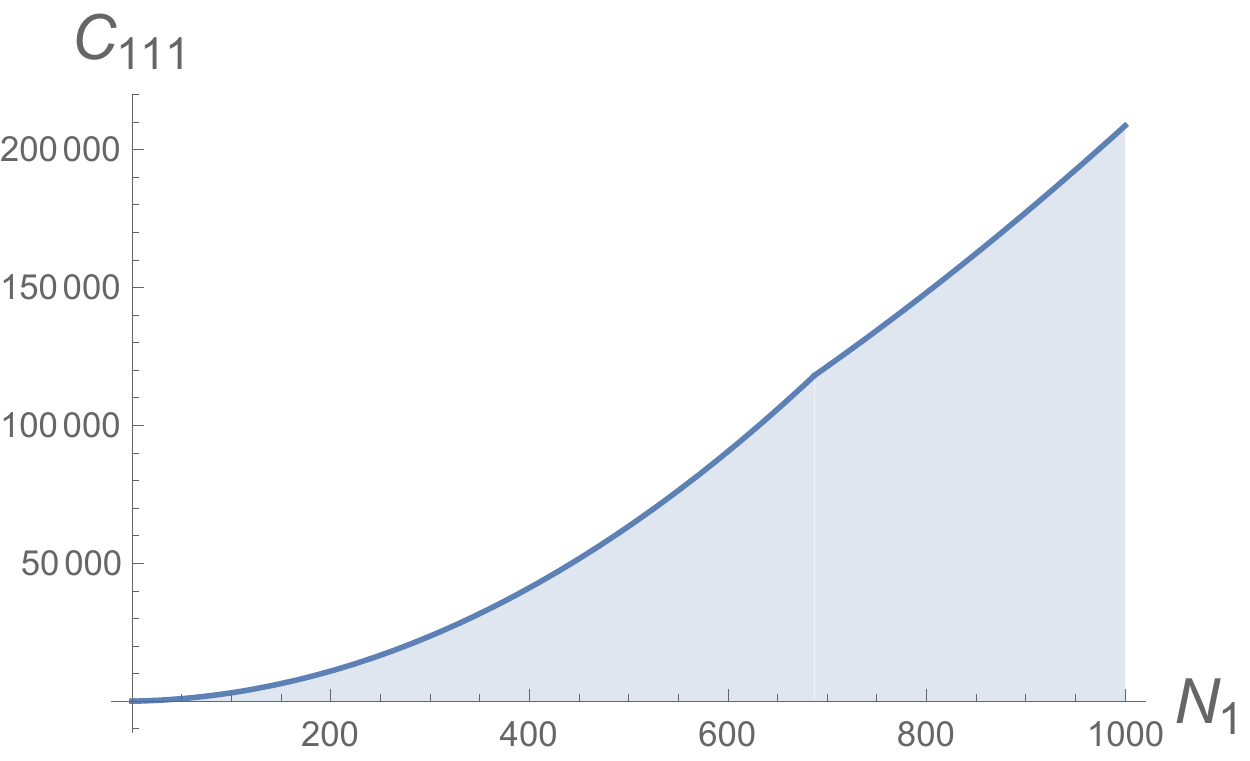}
		\caption{Upper bound on $C_{111}$ for $c=40$.}
		\label{c40_1}
	\end{minipage}\hfill
	\begin{minipage}{0.5\textwidth}
		\centering
		\includegraphics[width=1\textwidth]{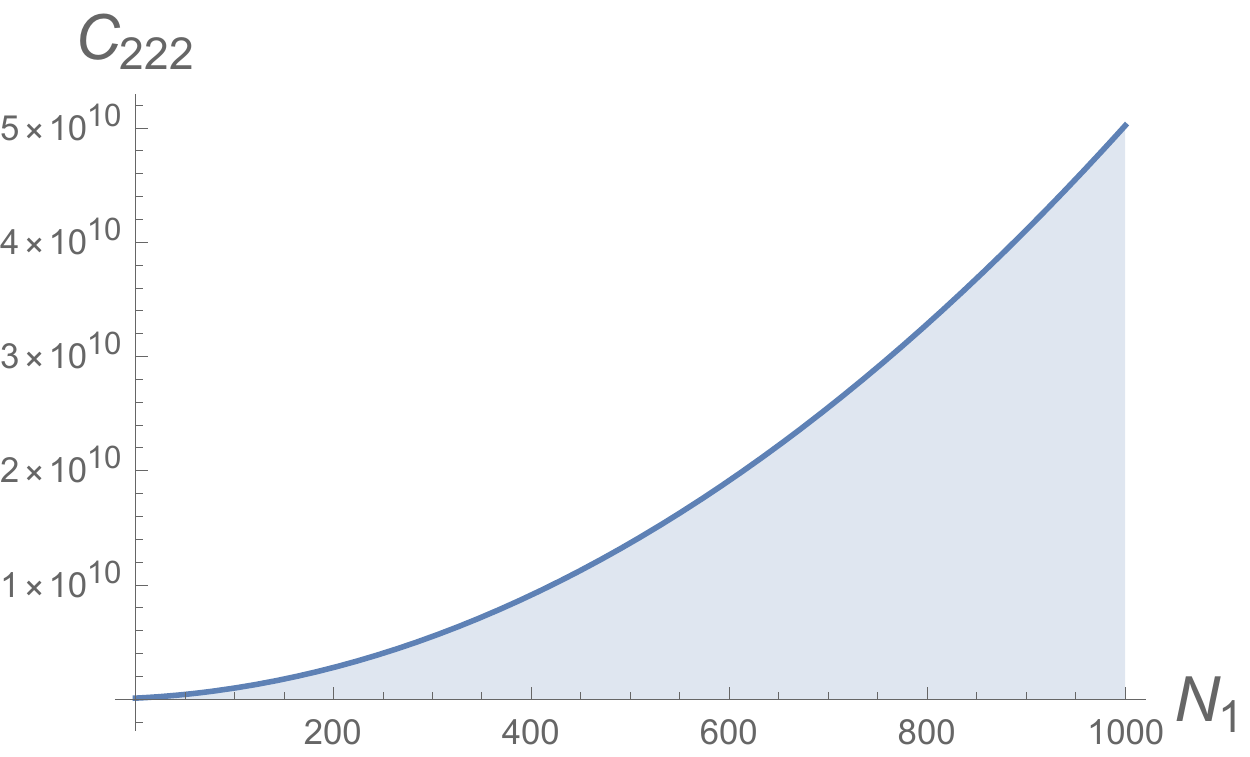}
		\caption{Upper bound on $C_{222}$ for $c=40$.}
		\label{c40_2}
	\end{minipage}
\end{figure}

To give a better picture,
we can also derive upper bounds on $C_{111}$ and $C_{222}$ as a function of $N_1$ only. To do this first note that coefficients of the form $C_{1h_2h_3}$  depend only on $C_{111}$ and $N_1$. For instance, we have
\be \label{c40C112}
C_{112} = -\frac{C_{111}}{2}+N_1^2+\frac{19 N_1}{20},
\ee
which directly gives an upper bound
\be
C_{111}\leq	2N_1\left(N_1+\frac{19}{20}\right).
\ee
Similarly, higher $C_{1h_2h_3}$ give an upper bound on $C_{111}$ which is piecewise quadratic in $N_1$, and is plotted in figure~\ref{c40_1}. This allows us to find a global bound for the average square of the three point function $c_{111}$ by maximising over $N_1$:
\be
\langle c_{111}^2\rangle=\frac{C_{111}}{N_1^3} \leq \frac{39}{10}\ .
\ee
To find similar bounds on $C_{222}$, we consider higher coefficients $C_{h_1h_2h_3}$ with $h_i>1$, which depend on all three parameters of the light data, \emph{i.e.}, $N_1$, $C_{111}$ and $C_{222}$. We eliminate the dependence on $C_{111}$ using the bound obtained for $C_{111}$ in terms of $N_1$. In this way, we derive an upper bound on $C_{222}$, which we have plotted in figure~\ref{c40_2}. The averaged quantity $\langle c_{222}^2\rangle$ has the following global bound:
\be
\qquad \langle c_{222}^2\rangle \lesssim 9.48\cdot 10^{-6}\ .
\ee

\subsection{c=48}

The multiplicities of primary fields with higher dimensions are given in terms of $N_1$ and $N_2$. For the first few multiplicities we have
\bea\label{c48mult}
&& N_3= 196882 N_1 - N_2 + 42987519,\qquad\quad\; N_4 = 15625 (1363 N_1+2588731) - N_2,\nonumber\\
&&N_5 =  2773 (303862 N_1+3052113849),\qquad N_6=4189 ( 4621643 N_1+189648425948),\nonumber\\
&&N_7 =312092484375 N_1 + N_2+44323300777781250,\nonumber\\
&&N_8= 3875 (1006083871 N_1+439403786287702).
\eea
We observe that some of the expressions depend on both $N_1$ and $N_2$, while others depend only on $N_1$. This is in general the case for theories with central charge $c=24\ell$, $\ell\in\mathbb Z$: the torus partition function of such theories contains a term which is proportional  to $\Delta^{c/24}=\Delta^\ell$, where $\Delta$ is the discriminant (\ref{disc}). Taking into account the pre-factor of the character (\ref{g1char}) which is of the form $ {q / \Delta} ^ {c/24} $, we find that the contribution of this term to the character is only one term: $a_{\ell}\,q^{\ell}$, where $a_{\ell}$ is a constant (see equations (\ref{Wg1c48_ii}) and (\ref{Wg1c72_i})). The coefficient of each term in the $q$-expansion of the character ($a_h\,q^h$) includes the contribution from Virasoro primaries of that dimension (\emph{i.e.}, $N_h$) as well as Virasoro descendants of primary fields of lower dimensions:
\be\label{Nh}
a_h=N_h+\sum_{i=0}^{h-1}p(h-i)\,N_i,
\ee
where $p(h-i)$ are the integer partitions corresponding to descendants at level $(h-i)$.{\footnote{We note that since $L_{-1}$ annihilates the vacuum, the counting of the number of descendants is different for $i=0$ compared to $i>1$.}} Thus, the contribution of $a_\ell$ to $N_h$, $h>\ell$, only appears in the descendants of lower dimensional primaries and so is dressed with the integer partitions corresponding to the descendant level. It is then easy to see that, for some values of $h>\ell$, the cancellations between the terms containing $a_\ell$ (dressed with integer partitions) yields vanishing of $N_\ell$ in the expression for $N_h$, as observed above in equation (\ref{c48mult}) above for $\ell=2$ (see also (\ref{c72mult}) for the case of $\ell=3$). We note that this is only the case for theories with $c=24 \ell$,  and suggests that in these theories the multiplicity of the `heavy' field with $h=c/24$ either does not affect the number of heavier primary fields ($N_{h}$, $h> c/24$), or comes in with a coefficient of $O(1)$ whereas the coefficients of $N_{h_i}$ with $h_i < c/24$ are orders of magnitude larger.

Imposing (\ref{Nh_bound}) and (\ref{3pf_bound}), we then obtain the following bounds on $N_2$, which is valid if $N_1>0$:
\be\label{c48_N2_bound}
23 N_1 + 71 \le N_2 \leq \frac15(78739 N_1 + 11951203)\ .
\ee	
Here the upper and lower bound come from $C_{133}$ and $C_{122}$, respectively.

\begin{figure*}
    \centering
    \begin{minipage}[b]{.45\linewidth}
        \centering
        \includegraphics[width=7cm]{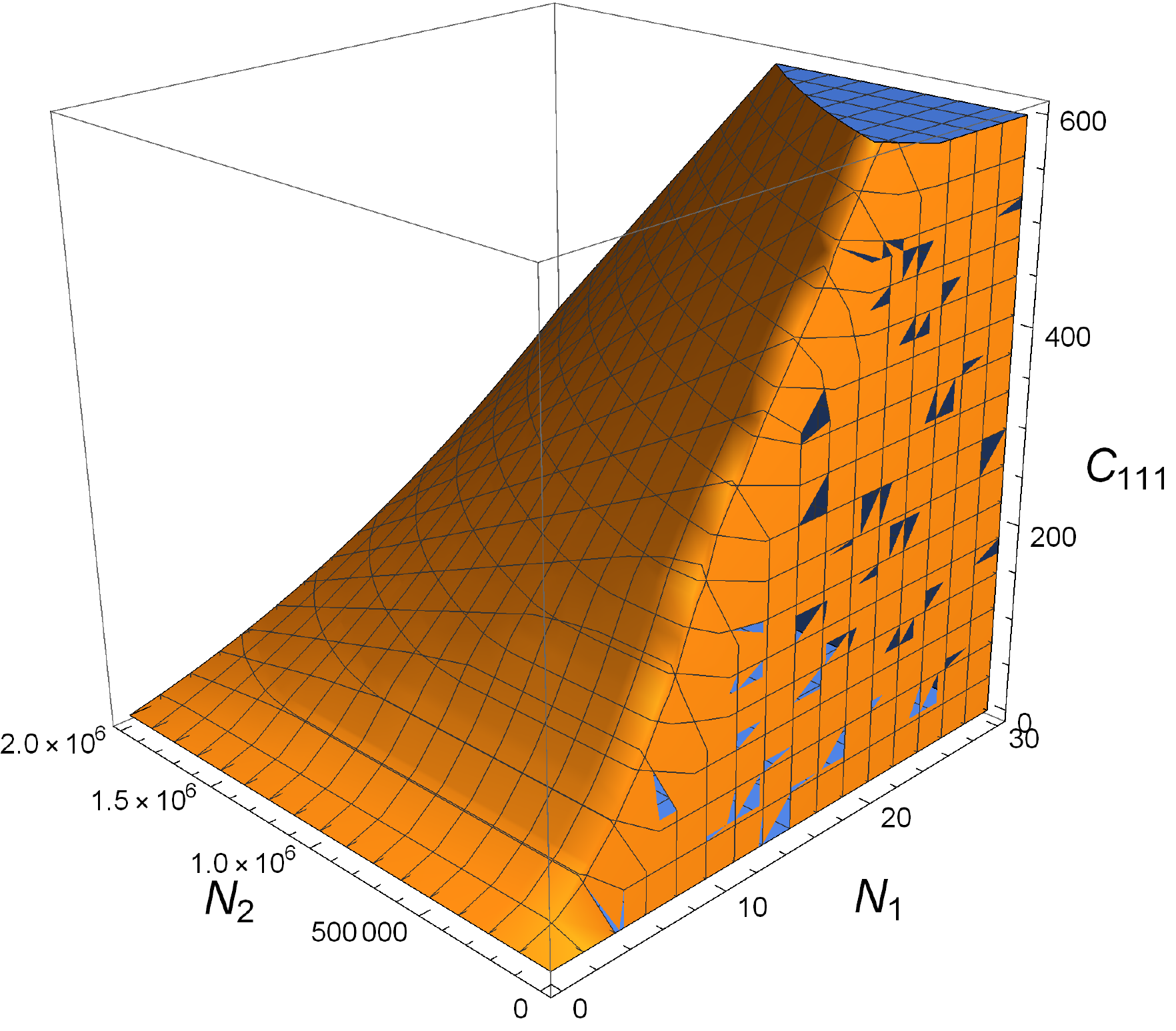}
    \end{minipage}%
    \hfill%
    \begin{minipage}[b]{.45\linewidth}
    \centering
        \centering
        \includegraphics[width=7cm]{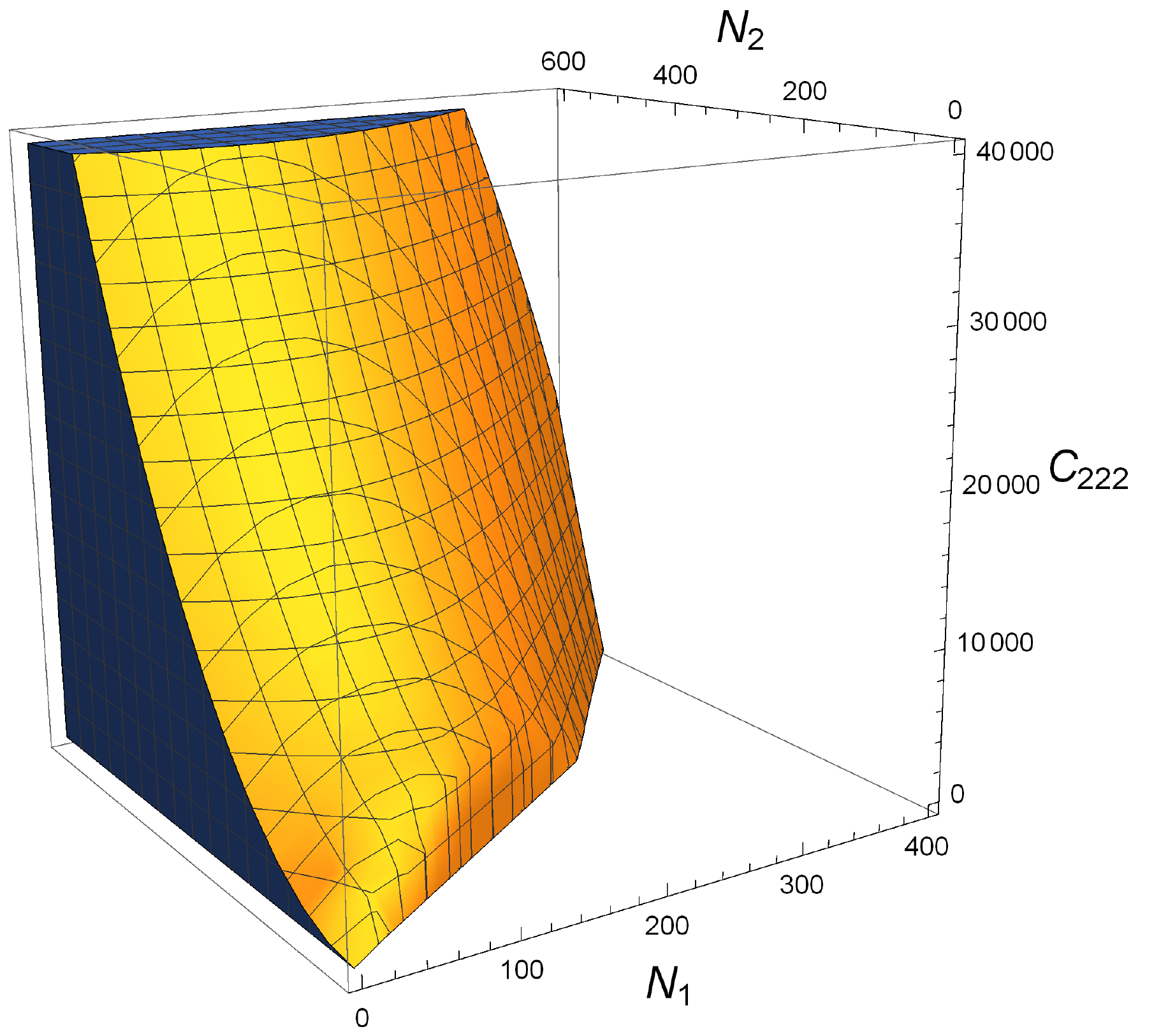}
    \end{minipage}\\[-7pt]
    \begin{minipage}[t]{.45\linewidth}
        \caption{Parameter space of $N_1$, $N_2$, and $C_{111}$ for CFTs with $c=48$ satisfying (\ref{Nh_bound}) and (\ref{3pf_bound}).}
       \label{fig_c48_C111}
    \end{minipage}%
    \hfill%
    \begin{minipage}[t]{.45\linewidth}
        \caption{Parameter space of $N_1$, $N_2$, and $C_{222}$ for CFTs with $c=48$ satisfying (\ref{Nh_bound}) and (\ref{3pf_bound}).}
        \label{fig_c48_C222}
    \end{minipage}%
\end{figure*}

Next we obtain a bound on $C_{111}$ by considering the coefficients $C_{1h_2h_3}$, which again only depend on $N_1$, $N_2$, and $C_{111}$, yielding
\be\label{c48_C111_bound}
0 \le C_{111} \leq \frac{504}{3029}\,N_1 (N_1+152 ),
\ee
where the upper bound comes from $C_{133}$, and we have used (\ref{c48_N2_bound}) to eliminate the dependence on $N_2$. If we want to keep the dependence on $N_2$, the result is shown in figure \ref{fig_c48_C111}.

Finally, by considering the higher coefficients $C_{h_1h_2h_3}$, and using  (\ref{c48_C111_bound}) to bound $C_{111}$, we find the region in the space of parameters $N_1$, and $N_2$, and $C_{222}$ which satisfies all the constraints. This region is depicted in figure \ref{fig_c48_C222}.

We note that the $c=48$ extremal CFT partition function, for which we have $N_1=N_2=C_{111}=C_{222}=0$, satisfies all our inequalities and is compatible with our results.

\subsection{c=56}
\begin{figure}[t]
	\begin{center}
		\includegraphics[scale=0.5]{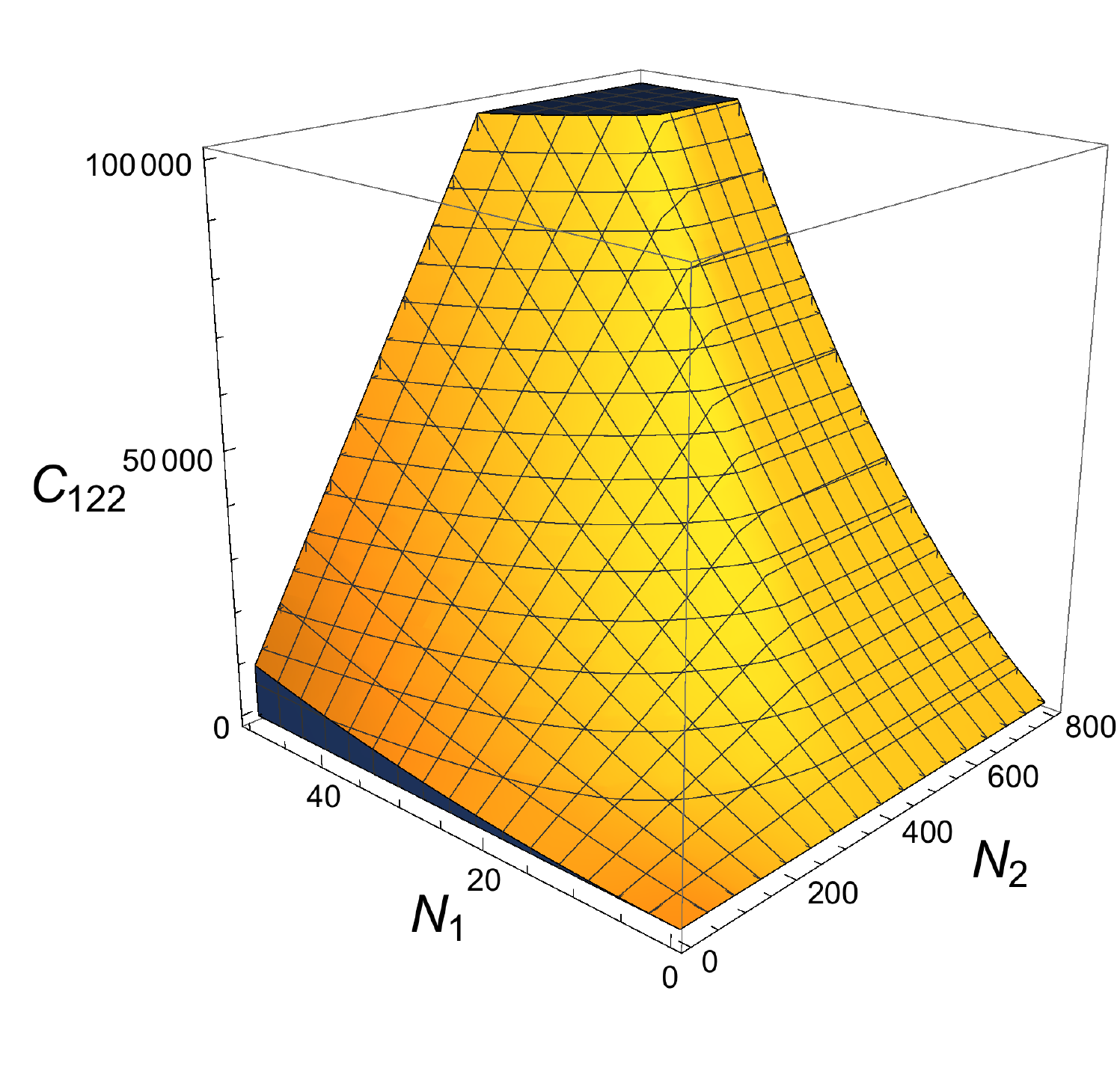}
        \caption{Allowed parameter space of $N_1$, $N_2$, and $C_{122}$ for CFTs with $c=56$.}
        \label{c56C212}
	\end{center}
\end{figure}

The multiplicities $N_h$ are fixed by $N_1$ and $N_2$ as
\bea\label{c56mult}
&& N_3 = 139750 N_1 + 247 N_2 + 7402775,\\
&& N_4 = 851 (81313 N_1+ 39875487) + 3875 N_2,\nonumber\\
&& N_5 = 5679389 (1220 N_1+ 2985119) + 30380 N_2.\nonumber
\eea
The coefficients $C_{112}$ depends on $N_1$ and $C_{111}$, and yields the following bound on $C_{111}$
\be 
C_{111}\leq \frac{1}{14} \left(28 N_1^2+27 N_1\right).
\label{c56_C111_bound}
\ee
Similarly, we can find bounds on the other light data. As an example, figure \ref{c56C212} shows the region in the parameter space  where the constraints (\ref{Nh_bound}) and (\ref{3pf_bound}) are satisfied for $C_{122}$.

\subsection{c=64}
The multiplicities of primary fields depend on $N_1$ and $N_2$. For the first few low lying values we have
\bea\label{c64mult}
&& N_3 = 3(7038 N_1 + 165 N_2 + 92837),\\
&& N_4 = 86714875 N_1 + 69255 N_2 + 13996384631,\nonumber\\
&& N_5 =  140095 (602550 N_1+ 483860983) + 2044760 N_2.\nonumber
\eea
From $C_{112}$ we find the bound 
\be 
C_{111}\leq \frac{1}{16} \left(32 N_1^2+31 N_1\right).
\label{c64_C111_bound}
\ee
As an example for the other light data, we again plot the allowed region for $C_{122}$ in figure \ref{c56C212}.

\begin{figure}[t]
	\begin{center}
		\includegraphics[scale=0.6]{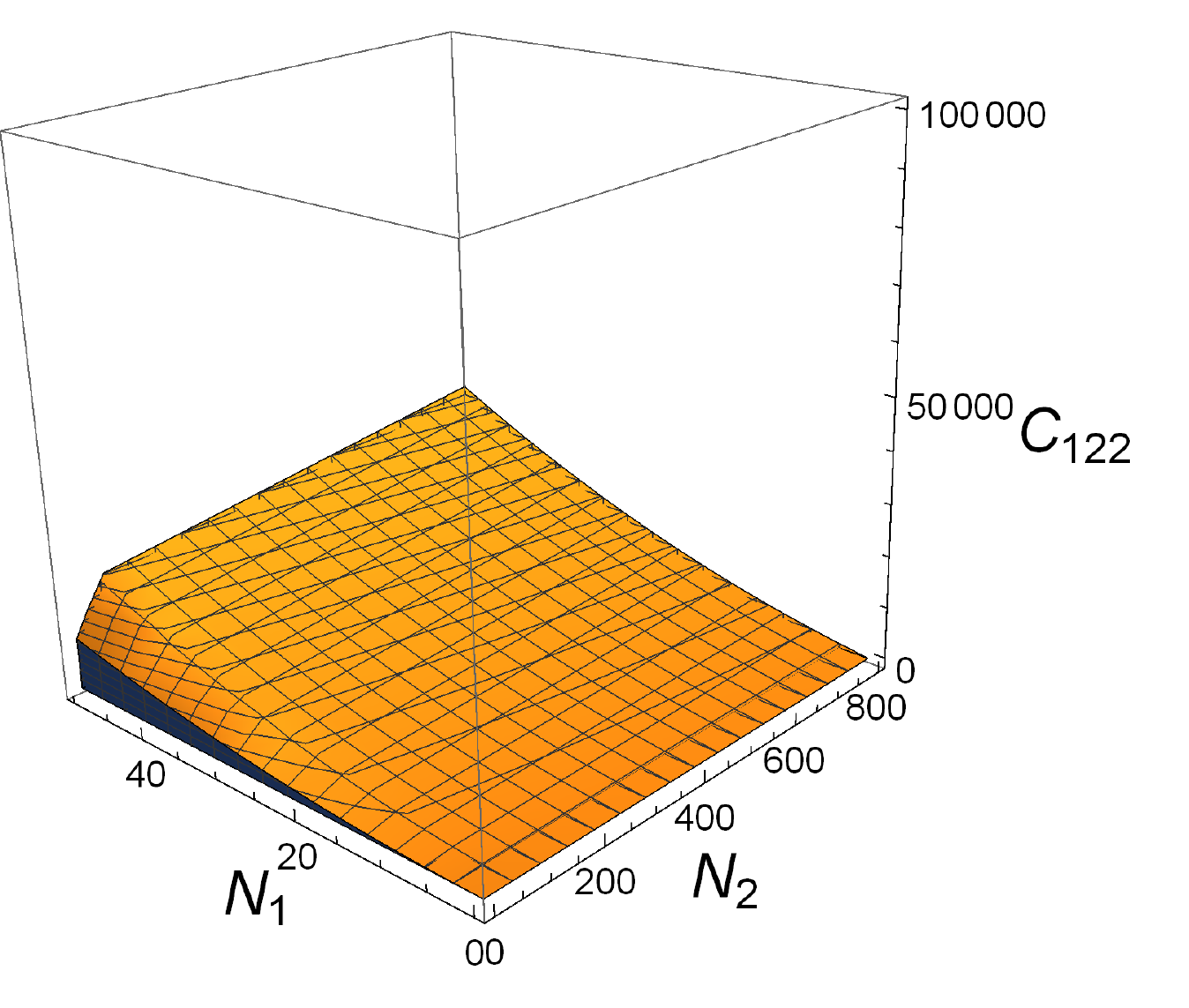}
		\caption{Allowed parameter space of $N_1$, $N_2$, and $C_{122}$ for CFTs with $c=64$.}
		\label{c64C212}
	\end{center}
\end{figure}

\subsection{c=72}
The multiplicities of primary fields $N_h$ are determined by $N_1$, $N_2$, and $N_3$, and for the first few values of $h$ we have
\bea\label{c72mult}
&& N_4= 43184401 N_1+196882 N_2-N_3+2593096792,\\
&&N_5 = 40470218750 N_1+21296875 N_2-N_3+12753498297254,\nonumber\\
&&N_6 =  8464354312603 N_1+842609326 N_2+9516562869359272,\nonumber\\
&&N_7 =71 (11189529807869 N_1+272676937 N_2+36808438051520256).\nonumber
\eea
Again we see that some of the multiplicities  only depend on $N_1$ and $N_2$ and not on $N_3$. If $N_1>0$, then from $C_{144}$ and $C_{133}$ we obtain upper and lower bounds on $N_3$:
\be\label{c72_N2_bound}
94 N_1+23 N_2+119 \le N_3 \leq \frac15({12029942 N_1}+{78739 N_2}+{585513523})\ .
\ee	
Note that again the extremal partition function satisfies all our constraints and is thus compatible with unitarity.

\section*{Acknowledgments}
This work is partly based on the master thesis of one of us (GM). We would like to thank Matthias Gaberdiel and Peter Goddard for sharing a draft of their upcoming book with us, and Roberto Volpato for sharing his Mathematica notebooks with us.  CAK thanks the Harvard University High Energy Theory Group for hospitality. IGZ thanks Institute of Physics at the University of Amsterdam for hospitality. We thank the Galileo Galilei Institute for Theoretical Physics (GGI) for the hospitality and INFN for partial support during the completion of this work, within the program \textit{New Developments in AdS$_3$/CFT$_2$ Holography}. In addition, IGZ thanks INFN as well as the ACRI (Associazione di Fondazioni e di Casse di Risparmio S.p.a.) for partial support through a YITP fellowship. CAK and IGZ are supported by the Swiss National Science Foundation through the NCCR SwissMAP.

\appendix

\section{Eisenstein series of degree one and two}\label{s_eisen}
The Eisenstein series of degree $g$ and weight $k$ are defined as
\be\label{eisen}
E^{(g)}_{k}(\Omega)=\sum_{C,D}\mathrm{det}(C\Omega+D)^{-k},\qquad
\bigg(\!\!\begin{array}{cc}A & B\\C & D \end{array}\!\!\bigg)\in\mathrm{Sp}(2g,\mathbb Z),
\ee

\subsection{Degree one}
The Eisenstein series of degree one are of the form
\be\label{eisen_g1}
E^{(1)}_{k}(\tau)=\sum_{\substack{c,d\in\mathbb Z\\(c,d)\ne(0,0)}}\frac1{(c\tau+d)^{k}},\qquad
\bigg(\!\!\begin{array}{cc}a & b\\c & d \end{array}\!\!\bigg)\in\mathrm{SL}(2,\mathbb Z).
\ee
The Eisenstein series have Fourier expansions in terms of the multiplicative period $q=e^{2\pi i\tau}$. It is customary to use the normalisation
\be\label{eisen_g1_four_ii}
G_{k}(\tau):=\frac{E^{(1)}_{k}(\tau)}{2\zeta(k)}=1+\frac{2}{\zeta(1-k)}\sum_{n=1}^\infty\sigma_{k-1}(n)\,q^n\ ,
\ee
where $\zeta(k)$ is the Riemann $\zeta$ function
and $\sigma_{k}(n)$ is the divisor sum function
\be\label{div_i}
\sigma_{k}(n)=\sum_{d|n}d^k.
\ee

\subsection{Degree two}\label{app_g2}
The Fourier expansion of the Eisenstein series of degree two in terms of the period matrices $\tau_{ij}$ is of the form \cite{vandergeer:2008}
\be\label{eisen_g2_four}
E^{(2)}_{k}(\Omega)\equiv E_{k}(\Omega)=\sum_Na_k(n,m,r)\,e^{2\pi i\,\mathrm{tr}(N\Omega)},
\ee
where $\Omega$ is given by (\ref{Omega_g2})
and
\be\label{N}
N=\Bigg(\!\!\begin{array}{cc}n&\frac r2\\\frac r2&m\end{array}\!\!\Bigg).
\ee
In terms of the multiplicative periods we have
\be\label{eisen_g2_four_ii}
E_{k}=\sum_{n,m=0}^\infty\sum_{\substack{r\in\mathbb Z\\4nm-r^2\ge0}}a_k(n,m,r)\,q_{11}^n\,q_{22}^m\,q_{12}^{r}\ .
\ee
The Fourier coefficients $a_k(n,m,r)$ are given by 

\be 
a_k(n,m,r) = \frac{2}{\zeta(3-2k)\zeta(1-k)}\sum\limits_{d|(n,m,r)}d^{k-1}H\left(k-1,\frac{4nm-r^2}{d^2}\right),
\ee
where $H(k-1,N)$ is the Cohen function introduced in \cite{Cohen1975}. It can be computed from a modified version of the Dirichlet L-series as $H(k-1,N) = L_{-N}(2-k)$ (page 21 of \cite{MR781735}). This modified Dirichlet L-series in turn is defined as
\be 
L_D(s) = \left\{\begin{array}{ll}
	0       & \text{if } D\neq 0,1 \text{   (mod4)}, \\
	\zeta(2s-1) & \text{if } D=0,\\
	L_{D_0}(s)\sum\limits_{d|f}\mu(d)\left(\frac{D_0}{d}\right)d^{-s}\sigma_{1-2s}(f/d) & \text{if } D=0,1 \text{   (mod4) , $D\neq 0$}.
\end{array}\right.
\label{modifiedDirichlet}
\ee
Here $D_0$ is the discriminant of the algebraic number field $\mathbb{Q}(\sqrt{D})$, and $f$ is given by $D=D_0 f^2$. $\left(\frac{D_0}{d}\right)$ is the Kronecker symbol, and  $L_{D_0}(s)$ is its associated Dirichlet L-series defined in the usual way as $L_{D_0}(s)= L\left(s,\left(\frac{D_0}{\cdot}\right)\right)= \sum_{n=1}^{\infty}\left(\frac{D_0}{n}\right)n^{-s}$. $\mu$ is the M\"obius $\mu$-function, and $\sigma$ is again the divisor sum function (\ref{div_i}). The Cohen function is always rational with a bounded denominator and a table with some values of the Cohen function is provided in \cite{Cohen1975}. We choose the normalization such that $a_k(0,0,0)=1$.

The other two generators of the ring of degree 2 modular forms (\ref{g2gens}) can be written as
\be\label{chi10}
\chi_{10}=\frac{43867}{2^{12}\cdot3^5\cdot5^2\cdot7\cdot53}(E_{10}-E_4E_6)\ ,
\ee
and
\be\label{chi12}
\chi_{12}=\frac{131\cdot593}{2^{13}\cdot3^7\cdot5^3\cdot7^2\cdot337}(3^2\cdot7^2E^3_{4}+2\cdot5^3E_6^2-691-E_{12})\ .
\ee

\section{More on genus two partition functions}\label{app_Z2}
This appendix contains the genus 1 and genus 2 partition functions for meromorphic CFTs with central charge $c=56$, $c=64$, and $c=72$.

\subsection{c=56}\label{app_c56Z2}
The light data of the CFT has 6 elements: $\cL=\{N_1, N_2, C_{111}, C_{122}, C_{222}, C_{223}\}$. We find
\be\label{Wg1c56_ii}
W^{c=56}_1=G_4^7+(N_1-1736)G_4^4\,\Delta+(-991N_1+N_2+401661)G_4\,\Delta^2.
\ee
The character then reads
\bea\label{g1charc56}
&&\chi_{1}^{c=56}=q^2\Big(j(\tau)^2+(N_1-1736)j(\tau)+(N_2-991N_1+401661)\Big)\chi^{c=8}_1\nonumber\\
&&\qquad\;\;=1+N_1q+(1+N_1+N_2)q^2 +(7402776+139752N_1+248N_2)q^3+\nonumber\\
&&\qquad\qquad\;+\,(4124 N_2+69337116 N_1+33941442214) q^4+O(q^5).
\eea

There are 10 terms contributing to the genus 2 amplitude:
\begin{align}\label{Z_g2_c56}
Z_2^{c=56} = \frac{1}{(F_{12})^{7/3}}&\Big(b_1\,E_4^7 + b_2\,E_4^4\,\psi_{12}+b_3\,E_4^4\,\chi_{12}+b_4\,E_4\,\psi_{12}^2+b_5\, E_4\,\chi_{12}^2+b_6\, E_4\,\psi_{12}\,\chi_{12}+\nonumber\\
&\!\!\!+b_7\,E_4^2\,\chi_{10}^2+b_8\,E_6^3\,\chi_{10}+b_9\,E_6\,\psi_{12}\,\chi_{10}+b_{10}\,E_6\chi_{12}\,\chi_{10}\Big),
\end{align}
where
\begin{align}\label{Z_g2_c56_i}
&b_1=1,\qquad b_2=N_1-1736,\qquad b_3=N_1 (N_1+238)-2 (N_2+394717),\\
&b_4=N_2-991N_1+401661,\qquad\; b_5=(-991 N_1+N_2+401661) (737 N_1+N_2+387837),\nonumber\\
&b_6=(-N_1-1720) (991 N_1-N_2-401661),\nonumber\\
&b_7= \frac{1}{1827}\Big(137030075 N_1^2+114086 N_2 N_1+35003496654 N_1+203  N_2^2-203522406 N_2+\nonumber\\   
&\qquad \qquad +57549384551511 -1781149608 C_{111}-2322656 C_{122}+26796 C_{222}-2436 C_{223}\Big),\nonumber\\
&b_8=\frac{1}{3} \left(N_1^2-26 N_1-2 N_2+918790-12 C_{111}\right),\nonumber\\
&b_9= \frac{1}{3} \left(1457 N_1^2+N_2 N_1+264101 N_1-3704 N_2+1488057192-17772 C_{111}-12 C_{122}\right),\nonumber\\
&b_{10}=\frac{2}{609} (-15219925 N_1^2+75110 N_2 N_1+30968282142 N_1+203 N_2^2-20201052 N_2\nonumber\\
&\qquad\qquad\;-12792520067067+445554144 C_{111}-1504804 C_{122}-1218 C_{222}-1218 C_{223}).\nonumber
\end{align}

\subsection{c=64}\label{app_c64Z2}
The light data of the theory has 8 elements in this case: $\cL=\{N_1, N_2, C_{111}, C_{122}, C_{222}$, $C_{223}, , C_{233}, C_{333}\}$. The degree 1 modular form $W^{c=64}_1$ is found to be
\be\label{Wg1c64_i}
W^{c=64}_1=G_4^8+(N_1-1984)G_4^5\,\Delta+(N_2-1239 N_1+705057)G_4^2\,\Delta^2,
\ee
and the character reads
\bea\label{g1charc64}
&&\chi_{1}^{c=64}=q^2\Big(j(\tau)^2+(N_1-1984)j(\tau)+(N_2-1239 N_1+705057)\Big)(\chi^{c=8}_1)^2\nonumber\\
&&\qquad\;\;=1+N_1q+(1+N_1+N_2)q^2 +4(124 N_2+5279 N_1+69628)q^3+\nonumber\\
&&\qquad\qquad\;+\,8\,(8719 N_2+10841999 N_1+1749582893) q^4+O(q^5).
\eea

The genus 2 partition function consists of contributions from 12 terms:
\begin{align}\label{Z_g2_c64}
Z_2^{c=64} =& \frac{1}{(F_{12})^{8/3}}\Big(b_1\, E_4^8 + b_2\, E_4^5\,\psi_{12}+b_3\, E_4^5\,\chi_{12}+b_4\, E_4^2\,\psi_{12}^2 + b_5\,E_4^2\,\chi_{12}^2+ b_6\, E_4^2\,\psi_{12}\,\chi_{12} +\nonumber\\
&\qquad\;\;\,+\,b_7\, E_4^4\,E_6\,\chi_{10}+b_8\, E_4^3\,\chi_{10}^2+b_9\,E_4\,E_6\,\psi_{12}\chi_{10}+b_{10}\,E_4\,E_6\,\chi_{12}\,\chi_{10},\nonumber\\
&\qquad\;\;\,+ b_{11}\,\psi_{12}\,\chi_{10}^2+b_{12}\,\chi_{12}\,\chi_{10}^2\Big),
\end{align}
where
\begin{align}\label{Z_g2_c64_i}
&b_1=1,\qquad b_2=N_1-1984,\qquad b_3=N_1 (N_1+238)-2 (N_2+451105),\\
&b_4=N_2-1239 N_1+705057 ,\quad\;\;\, b_5=(-1239 N_1+N_2+705057)(489 N_1+N_2+262689),\nonumber\\
&b_6=(-N_1-1472)(1239 N_1-N_2-705057) ,\nonumber\\
&b_7= \frac{1}{3} (N_1^2-26 N_1-2 N_2+1050046-12 C_{111}),\nonumber\\   
&b_8=\frac{1}{198} (15101526 N_1^2+1452 N_2 N_1+3334984620 N_1+22 N_2^2-22587411 N_2+\nonumber\\
&\qquad\quad+\,7475259339009-192900048 C_{111}-120768 C_{122}+2904 C_{222}-264 C_{223}),\nonumber
\end{align} 
\begin{align}
&b_9= \frac{1}{3} (-519 N_1^2+N_2 N_1+802641 N_1-496 N_2-349708272+5940 C_{111}-12 C_{122}),\nonumber\\
&b_{10}=\frac{1}{66} (-309012 N_1^2-5544 N_2 N_1+11166410904 N_1+44N_2^2-3344715 N_2+\nonumber\\
&\qquad\qquad-7826350562613+161128704 C_{111}-195216 C_{122}-264 C_{222}-264 C_{223}),\nonumber\\
&b_{11}=\frac{1}{1089} (-24762459808 N_1^2+92107488 N_2 N_1+25689118498597 N_1-243936 N_2^2+\nonumber\\
&\qquad\qquad-22372632711 N_2-9067003012774080+306936877136 C_{111}+\nonumber\\
&\qquad\qquad-195216 C_{212}-264 C_{222}-264 C_{223}),\nonumber\\
&b_{12}=\frac{4}{363} (-94717727296 N_1^2+2931120192 N_2 N_1+589970738884957 N_1-7248384 N_2^2+\nonumber\\
&\qquad\qquad-723196407978N_2-425511987841071945+20051460666080 C_{111}+\nonumber\\
&\qquad\qquad-27307245504 C_{212}-1034989956 C_{222}+24772572 C_{223}+435600 C_{323}+17424C_{333}).\nonumber
\end{align}

\subsection{c=72}\label{app_c72Z2}
The light data of the theory is $\cL=\{N_1, N_2, N_3,C_{111}, C_{122}, C_{222}, C_{223}, C_{233}, C_{333}, C_{334}\}$. The modular form $W^{c=72}_1$ contains contributions from four terms:
\bea\label{Wg1c72_i}
&&W^{c=72}_1=G_4^9+(N_1-2232)G_4^6\Delta+(-1487 N_1+ N_2+1069957)G_4^3\Delta^2+\nonumber\\
&&\qquad\qquad\quad\;\,+\,(159026 N_1 - 743 N_2 + N_3 - 36867719)\Delta^3,
\eea
and, using (\ref{g1char}), the genus 1 partition function is found to be
\bea\label{g1charc72}
&&\chi_{1}^{c=72}=1+N_1q+(1+N_1+N_2)q^2 +(1+2N_1+N_2+N_3)q^3+\\
&&\qquad\qquad\;\,+\,\Big(43184404 N_1+ 54 (3646 N_2+48020311)\Big) q^4+\nonumber\\
&&\qquad\qquad\;\,+\,4\,\Big(10128350789 N_1 + 512 (10495 N_2+6228560251)\Big)q^5+\nonumber\\
&&\qquad\qquad\;\,+\,54\,\Big(157498350003 N_1+16005555 N_2+176468917663891\Big)q^6 + O(q^7).\nonumber
\eea

The genus 2 amplitude consists of 17 terms of the form:
\bea\label{Z_g2_c72}
Z_2^{c=72}=&&\!\!\!\!\!\!\!\!\frac1{\left(F^{12}_2\right)^{3}}\bigg(b_1\,E_4^9+b_2\,E_4^6\,\psi_{12}+b_3\,E_4^6\,\chi_{12}+b_4\,E_4^3\,\psi_{12}^2+b_5\,E_4^3\chi_{12}^2+b_6\,E_4^3\,\psi_{12}\,\chi_{12}+\nonumber\\
&&\quad\;\,+\,b_7\,\psi_{12}^3+b_8\,\psi_{12}^2\,\chi_{12}+b_9\,\psi_{12}\,\chi_{12}^2+b_{10}\,\chi_{12}^3+b_{11}\,E_4^5\,E_6\,\chi_{10}+\nonumber\\
&&\quad\;\,+\,b_{12}\,E_4^4\,\chi_{10}^2+b_{13}\,E_4^2\,E_6\,\psi_12\,\chi_{10}+b_{14}\,E_4^2\,E_6\,\chi_12\,\chi_{10}+\nonumber\\
&&\quad\;\,+\,b_{15}\,E_4\,\psi_{12}\,\chi_{10}^2+b_{16}\,E_4\,\chi_{12}\,\chi_{10}+b_{17}\,E_6\,\chi_{10}^3\bigg),
\eea
where
\begin{align}\label{Z_g2_c72_i}
&b_1=1,\qquad b_2=N_1-2232,\\
&b_3=N_1^2+238 N_1-2 \left(N_2+507493\right),\qquad b_4=-1487 N_1+N_2+1069957,\nonumber\\
&b_5=-676419 N_1^2+\left(240 N_2-2 \left(N_3+39441780\right)\right) N_1+N_2^2+1803962 N_2+\nonumber\\
&\qquad-720 N_3+239514348745,\nonumber\\
&b_6=-1487 N_1^2+\left(N_2-1227209\right) N_1+3453 N_2-3 N_3+1420230525,\nonumber\\
&b_7=159026 N_1-743 N_2+N_3-36867719,\nonumber\\
&b_8=\left(N_1+2952\right) \left(159026 N_1-743 N_2+N_3-36867719\right),\nonumber\\
&b_9=\left(1969 N_1+N_2+2314117\right) \left(159026 N_1-743 N_2+N_3-36867719\right),\nonumber
\end{align}
\begin{align}
&b_{10}=91515328324 N_1^2+(-270936572 N_2+734500 N_3+27617614928260) N_1+\nonumber\\
&\qquad\;\,-\,731855 N_2^2+N_3^2+242 N_2 (N_3-1092878759)+270214322+N_3+\nonumber\\
&\qquad\;\,-\,11321414397534479,
\end{align}
\be
b_{11}=\frac{1}{3} \left(-12 C_{111}+N_1^2-26 N_1-2 N_2+1181302\right),
\ee
\bea
&b_{12}=\frac{1}{27}\,(2462547 N_1^2-1290 N_2 N_1+156350382 N_1+3 N_2^2-3746382 N_2+\\
&\quad\;\,+\,1296 N_3+1300027361735)-1137592C_{111}+\frac{17120 C_{122}}{333}+396 C_{222}-36 C_{223},\nonumber
\eea
\bea
&&b_{13}=\frac{1}{3} (-767 N_1^2+N_2 N_1+1003687 N_1+1485 N_2-3 N_3-685444851+\\
&&\qquad\quad\;\;+\,8916 C_{111}-12 C_{212}),\nonumber
\eea
\bea
&&b_{14}=-\frac{2}{9} (-74733 N_1^2-363 N_2 N_1+3 N_3 N_1-575177475 N_1-3 N_2^2-1013904 N_2+\nonumber\\
&&\qquad\qquad\;+\,1080 N_3+878731318367+18 C_{222}+18 C_{223})+\nonumber\\
&&\qquad\;+\,2927424 C_{111}-\frac{108088 C_{122}}{111},
\eea
\bea
&&b_{15}=-\frac{867016339094   N_1^2}{12321}+\frac{20152397 N_2 N_1}{111}-\frac{139 N_3 N_1}{9}+\frac{198581235265705 N_1}{4107}+\\
&&\qquad\;\,-\,\frac{1273 N_2^2}{9}+\frac{1146291334   N_2}{9}-\frac{N_2 N_3}{9}-\frac{2295505 N_3}{9}-\frac{256516494599113}{9}   \nonumber\\
&&\qquad\;\,+\,\frac{1158890673456 C_{111}}{1369}-\frac{190139008 C_{122}}{111}-14364 C_{222}+1348 C_{223}+4 C_{322}+16 C_{323},\nonumber
\eea
\bea
&&b_{16}=-\frac{12936962575678 N_1^2}{1369}+\frac{7297613773 N_2 N_1}{111}-\frac{10182755 N_3 N_1}{111}+\\
&&\qquad\;\,+\,\frac{23259488428568161
   N_1}{4107}+\frac{324899 N_2^2}{3}-\frac{N_3^2}{3}+\frac{94920984976 N_2}{3}-\frac{158 N_2 N_3}{3}+\nonumber\\
&&\qquad\;\,-\,\frac{30406417204 N_3}{573}-\frac{36705982837911919}{3}+\nonumber\\
&&\qquad\;\,+\,\frac{613750982045768 C_{111}}{1369}+\frac{20533968136 C_{122}}{111}-11890998 C_{222}+\nonumber\\
&&\qquad\;\,-\,170710 C_{223}-\frac{374664 C_{233}}{101}+24 C_{333}+16 C_{334},\nonumber
\eea
and,
\bea
&&b_{17}=\frac{103598772879382 N_1^2}{36963}-\frac{10744270517 N_2 N_1}{999}+\frac{2868763 N_3N_1}{999}+\\
&&\qquad\;\,-\,\frac{44331239306584169 N_1}{36963}+\frac{319738 N_2^2}{27}-\frac{2 N_3^2}{27}-\frac{137795278558 N_2}{27}+\nonumber\\
&&\qquad\;\,+\,\frac{632 N_2 N_3}{27}+\frac{50261856514 N_3}{5157}+\frac{17090981447176756}{27}+\nonumber\\
&&\qquad\;\,-\,\frac{143270294050616 C_{111}}{4107}+\frac{31421366264 C_{122}}{333}+1187070 C_{222}+\nonumber\\
&&\qquad\;\,-\,\frac{300838 C_{223}}{3}-\frac{286488 C_{233}}{101}-56
   C_{333}+\frac{16 C_{334}}{3}.\nonumber
\eea

\vspace{20pt}
\bibliographystyle{utphys}
\bibliography{ref}
\end{document}